\documentclass[12pt]{article}

\usepackage{epsfig,amsmath}

\title{\bf Electrovacuum Static  Counter-Rotating  Relativistic Dust Disks}

\author{Gonzalo Garc\'\i a R.\thanks{e-mail: gogarcia@uis.edu.co} \ and \ 
Guillermo A. Gonz\'{a}lez\thanks{e-mail: guillego@uis.edu.co}	\\
{\it Escuela de F\'{\i}sica, Universidad Industrial de Santander}	\\
{\it A.A. 678, Bucaramanga, Colombia}}

\begin{document}

\maketitle

\begin{abstract}

A detailed study of the Counter-Rotating Model (CRM) for generic  electros\-tatic  (magnetostatic) axially symmetric thin disks without radial pressure is presented. We find a general constraint over the counter-rotating tangential velocities needed to cast the surface energy-momentum tensor of the disk as the superposition of two counter-rotating charged  dust fluids.  We then show that this constraint is satisfied if we take the two counter-rotating     streams as circulating along electrogeodesics with  equal and opposite tangential velocities. We also find explicit expressions for the energy densities, electrostatic  (magnetostatic) charge densities and veloci\-ties  of the counter-rotating fluids.
Three specific examples are conside\-red  where we obtain some CRM well behaved based in simple solutions to the Einstein-Maxwell equations. The conside\-red solutions  are Reissner-Nordstr\"{o}m in the electrostatic case, 
its  magnetostatic counterpart  and two solutions obtained from  Taub-NUT and Kerr solutions.  
\vspace{0.4cm}

\end{abstract}

\section{Introduction}

Stationary or static axially symmetric exact solutions of Einstein  equations describing relativistic thin disks are of great astrophysical importance since can be used as models of certain stars, galaxies, accretion disk and universes. 
Theses  were first studied by Bonnor and Sackfield \cite{BS}, obtaining pressureless static disks, and then by Morgan and Morgan, obtaining static disks with and without radial pressure \cite{MM1,MM2}. In connection with gravitational collapse, disks were first studied by Chamorro, Gregory and Stewart \cite{CHGS}. Disks with radial tension have been also recently studied \cite{GL1}. Several classes of exact solutions
of the Einstein field equations corresponding to static and stationary
thin disks have been obtained by different authors [\ref{bib:LP} --
\ref{bib:GL2}], with or without radial pressure.

In the case of static disks without radial pressure, there are two common
interpretations. The stability of these models can be explained by either
assuming the existence of hoop stresses or that the particles on the disk
plane move under the action of their own gravitational field in such a
way that as many particles move clockwise as counterclockwise. This last
interpretation, the ``Counter-Rotating Model'' (CRM), is frequently made
since it can be invoked to mimic true rotational effects. Even though
this interpretation can be seen as a device, there are observational
evidence of disks made of streams of rotating and counter-rotating
matter \cite{RGK,RFF}.
 
Disklike sources in presence of electrogmatic fields, specially magnetic fields, are also of  astrophysical interest  mainly in the study of neutron stars, white dwarfs and galaxy formation. In the  context of general relativity  models of disks for Kerr-Newman metrics \cite{LBZ}, static axisymmetric spacetimes with magnetic fields \cite{LET1} and corformastationary metrics \cite{KBL}, have been considered recently. Following the Ref. \cite{LBZ} the resultating disks can also be interpreted either as rings with internal pressure and currents or as two counter-rotating  streams of freely moving charged particles, i.e. which move along electrogeodesics (solution to the geodesic equation in the presence of a Lorentz force). 

The   aim of this paper is to perform a detailed study of the CRM for generic  electros\-tatic  (magnetostatic) axially symmetric thin disks without radial pressure. The paper is organized as follows.  In Sec. 2 we present a summary of the procedure to obtain  thin disks models as rings with a purely azimuthal pressure  and currents using the well-known ``displace, cut and reflect'' method extended to solutions of Einstein-Maxwell equations.  In particular,   we obtain expressions for the surface energy-momentum tensor  and the electrostatic  (magnetostatic) current density of the disk. Next, in Sec. 3, the disks are interpreted in terms of the CRM. We find a general constraint over the counter-rotating tangential velocities needed to cast the surface energy-momentum tensor of the disk as the
superposition of two counter-rotating  charged dust fluids.  We then show that this constraint is satis\-fied if we take the two counter-rotating streams as circulating along electrogeodesics with  equal and opposite tangential velocities.  We also find explicit expressions for the energy densities,  electrostatic  (magnetostatic) current densities and velocities of the counter-rotating
fluids. In following section, Secs. 4,  three specific examples are considered based in simple solutions to  the Einstein-Maxwell equations. The considered solutions are Reissner-Nordstr\"{o}m in the electrostatic case, its  magnetos\-tatic counterpart,  and two solutions generated from  Taub-NUT and Kerr solutions. In particular, we study the tangential velocities, mass densities and  electros\-tatic (magnetostatic) charge  densities of both streams. Also the stability against radial perturbation is considered. Finally, in Sec. 5, we summarize our main results.

\section{ Static Relativistic Thin Disks}

In this section we present
a summary of the procedure to obtain electrostatic (magnetostatic) axially symmetric thin disks. The  simplest metric  to describe a  static axially symmetric spacetime is the Weyl's line element 
\begin{equation}
ds^2 = e^{- 2 \Phi} [r^2 d\varphi^2 + e^{2 \Lambda} (dr^2 + dz^2)]
\ - \ e^{2 \Phi} dt^2 \ , \label{eq:met}
\end{equation}
where $\Phi$ and $\Lambda$  are functions of $r$ and $z$
only.  The Einstein-Maxwell field equations, in  geometrized units such that $8 \pi G = c = \mu _0 = \varepsilon _0 = 1$,  are given by 
\begin{subequations}
\begin{eqnarray}
&   &    R_{ab} \  =  \ T_{ab},  \\
&   &     \nonumber       \\
&   &     T_{ab} \  =  \ F_{ac}F_b^{ \ c} - \frac 14 g_{ab}F_{cd}F^{cd},  \\
&   &     \nonumber    \\
&   &    F^{ab}_{ \ \ \ ; b} = 0,    \\
&   &     \nonumber       \\
&   &   F_{ab} =  A_{a,b} -  A_{b,a},
\end{eqnarray}\label{eq:einmax}
\end{subequations} 
where all symbols are understood.
For the metric (\ref{eq:met}), the Einstein-Maxwell equations in presense of purely electric field  are equivalent to the system
\begin{subequations}
\begin{eqnarray}
&    &  \Phi _{,rr} + \frac 1r  \Phi_{,r} +  \Phi_{,zz} - \frac {e ^{-2 \Phi}}{2} (\psi_{,r}^2
        + \psi_{,z}^2)      \ = \ 0,               \\
&    &     \nonumber          \\
&    &  \psi_{,rr} + \frac 1r \psi_{,r} + \psi_{,zz} - 2(\Phi_{,r} \psi_{,r} + \Phi_{,z} \psi_{,z} )  \ = \ 0,  \\
&    &     \nonumber \\
&    &   \Lambda _{,r} \ = \ r( \Phi_{,r}^2 - \Phi_{,z}^2) - \frac {re ^{-2 \Phi}}{2} (\psi_{,r}^2 - \psi_{,z}^2),  \\
&    &  \nonumber                                 \\
&    &   \Lambda _{,z} \ = \ 2r \Phi_{,r} \Phi_{,z} - r  e ^{-2 \Phi} \psi_{,r} \psi_{,z}, 
\end{eqnarray}\label{eq:ele}
\end{subequations}
and in the magnetostatic case to 
\begin{subequations}
\begin{eqnarray}
&    &  \Phi _{,rr} + \frac 1r  \Phi_{,r} +  \Phi_{,zz} - \frac {e ^{2 \Phi}}{2r^2} (A_{,r}^2 + A_{,z}^2)
        \ = \ 0,               \\
&    &     \nonumber          \\
&    &  A_{,rr} - \frac 1r A_{,r} + A_{,zz} + 2(A_{,r} \Phi_{,r} + A_{,z}\Phi_{,z} )  \ = \ 0,  \\
&    &     \nonumber \\
&    &    \Lambda _{,r} \ = \ r( \Phi_{,r}^2 - \Phi_{,z}^2) + \frac {e ^{2 \Phi}}{2r} (A_{,r}^2 - A_{,z}^2),  \\
&    &  \nonumber                                 \\
&    &  \Lambda _{,z} \ = \ 2r \Phi_{,r} \Phi_{,z} + \frac 1r  e ^{2 \Phi} A_{,r} A_{,z}, 
\end{eqnarray}\label{eq:mag}
\end{subequations}
where  $\psi$ and $A$ are the  magnetostatic  and electrostatic potencial, respectively, which  are also functions  of $r$ and $z$. 

In order to obtain a solution of (\ref{eq:mag}) - (\ref{eq:ele}) representing a thin disc at $z=0$, we assume that the components of the metric tensor are continuous across the disk, but their discontinuous first derivates on plane $z=0$, with discontinuity functions
$$
b_{ab} \ = g_{ab,z}|_{_{z = 0^+}} \ - \ g_{ab,z}|_{_{z = 0^-}} \ =
\ 2 \ g_{ab,z}|_{_{z = 0^+}} \ .
$$
Thus, the Einstein-Maxwell equations yield an energy-momentum tensor $T_a^b \
= \ Q_a^b \ \delta (z)$ and a planar current density ${\rm i}^a= 2F^{az} \delta (z) $ , where $\delta (z)$ is the usual Dirac function
with support on the disk and
$$
Q^a_b = \frac{1}{2}\{b^{az}\delta^z_b - b^{zz}\delta^a_b + g^{az}b^z_b -
g^{zz}b^a_b + b^c_c (g^{zz}\delta^a_b - g^{az}\delta^z_b)\}
$$
is the distributional energy-momentum tensor. The ``true'' surface
energy-momentum tensor (SEMT) of the disk, $S_a^b$, can be obtained
through the relation
\begin{equation}
S_a^b \ = \ \int T_a^b \ ds_n \ = \ e^{\Lambda -
\Phi} \ Q_a^b \ ,
\end{equation}
where $ds_n = \sqrt{g_{zz}} \ dz$ is the ``physical measure'' of length
in the direction normal to the disk, and the current density as ${\rm j}^a= \ e^{\Lambda -\Phi} {\rm i}^a $. For the metric (\ref{eq:met}), the  non-zero  components of  $S_a^b$ are
\begin{subequations}\begin{eqnarray}
&S^0_0 &= \ 2 e^{\Phi - \Lambda} \left\{ \Lambda,_z - \ 2 \Phi,_z  \
 \right\} , \label{eq:emt1}     		\\
&	&	\nonumber	\\
&S^1_1 &= \ 2 e^{\Phi - \Lambda} \Lambda,_z , \label{eq:emt2}
\end{eqnarray}\label{eq:emt}\end{subequations}
and   current density equal to
\begin{subequations}
\begin{eqnarray}
& {\rm j}_t &= \ -2 e^{\Phi - \Lambda} \psi _{,z} , \label{eq:corelec}   \\
&	&	\nonumber	\\
&{\rm j}_{\varphi} &= \ -2 e^{\Phi - \Lambda} A _{,z} , \label{eq:cormag} \end{eqnarray}\label{eq:cor}
\end{subequations}
in the electrostatic and magnetostatic cases, respectively. All the quantities are evalua\-ted at $z = 0^+$.

With an orthonormal tetrad ${{\rm e}_{\hat a}}^b = \{ V^b , W^b , X^b ,
Y^b \}$, where
\begin{subequations}\begin{eqnarray}
V^a &=& e^{- \Phi} \ ( 1, 0, 0, 0 ) \ ,	\\
	&	&	\nonumber	\\
W^a &=& \frac{e^\Phi} {r} \ \ ( 0, 1, 0, 0 ) \ ,	\\
	&	&	\nonumber	\\
X^a &=& e^{\Phi - \Lambda} ( 0, 0, 1, 0 ) \ ,	\\
	&	&	\nonumber	\\
Y^a &=& e^{\Phi - \Lambda} ( 0, 0, 0, 1 ) \ ,
\end{eqnarray}\label{eq:tetrad}\end{subequations}
we can write the metric and the SEMT in the canonical forms
\begin{subequations}\begin{eqnarray}
&	&g_{ab} \ = \ - V_a V_b + W_a W_b + X_a X_b + Y_a Y_b \ ,
\label{eq:metdia}							\\
&	&	\nonumber						\\
&	&S_{ab} \ = \ \epsilon V_a V_b + p_\varphi W_a W_b  \
, \label{eq:emtdia}
\end{eqnarray}\end{subequations}
where
\begin{equation}
\epsilon \ = \ - S^0_0 \quad , \quad p_\varphi \ = \ S^1_1 \quad  , \label{eq:dps}
\end{equation}
are, respectively, the energy density and the azimuthal pressure of the disk.

\section{The Counter-Rotating Model}

We now consider, based on references \cite{LET2} and \cite{FMP}, the
possibility that the SEMT $S^{ab}$ and the current density  ${\rm j}^a$ can be written as the superposition of
two counter-rotating  fluids that circulate in opposite directions;
that is, we assume 
\begin{subequations}
\begin{eqnarray}
S^{ab} &=& S_+^{ab} \ + \ S_-^{ab} \ , \label{eq:emtsum}   \\
       & &	        \nonumber	                    \\
{ \rm j}^a    &=& { \rm j}_+^a   +{\rm j}_-^a, \label {eq:corsum} 
\end{eqnarray}
\end{subequations}
where the  quantities in the right-hand side  are, respectively, the SEMT and the current density of the prograd and retrograd counter-rotating fluids. 

Let  $U_\pm^a = ( U_\pm^0 , U_\pm^1, 0 , 0 )$ be the velocity vectors
of the two counter-rotating fluids. In order to do the decomposition
(\ref{eq:emtsum}) and (\ref{eq:corsum}) we project the velocity vectors onto the tetrad ${{\rm
e}_{\hat a}}^b$, using the relations \cite{CHAN}
\begin{equation}
U_\pm^{\hat a} \ = \ {{\rm e}^{\hat a}}_b U_\pm^b \qquad , \qquad U_\pm^
a = \ U_\pm^{\hat c} {{\rm e}_{\hat c}}^a  .
\end{equation}
With the tetrad (\ref{eq:tetrad}) we can write
\begin{equation}
U_\pm^a \ = \ \frac{ V^a + {\rm U}_\pm W^a }{\sqrt{1 - {\rm U}_\pm^2}}
, \label{eq:vels}
\end{equation}
and thus
\begin{subequations}\begin{eqnarray}
&V^a &= \ \frac{\sqrt{1 - {\rm U}_-^2} {\rm U}_+ U_-^a - \sqrt{1 -
{\rm U}_+^2} {\rm U}_- U_+^a}{{\rm U}_+ - {\rm U}_-} \ , \label{eq:va} \\
&	&	\nonumber	\\
&W^a &= \ \frac{\sqrt{1 - {\rm U}_+^2} U_+^a - \sqrt{1 - {\rm U}_-^2}
U_-^a}{{\rm U}_+ - {\rm U}_-} \ , \label{eq:wa}
\end{eqnarray}\label{eq:vawa}\end{subequations}
where ${\rm U}_\pm = U_\pm^{\hat 1} / U_\pm^{\hat 0}$ are the tangential
velocities of the fluids with respect to the tetrad.

Using (\ref{eq:vawa}), we can write the SEMT as

\begin{eqnarray}
S^{ab} & = & \frac{ f( {\rm U}_- , {\rm U}_- ) (1 - {\rm U}_+^2) \
U_+^a U_+^b }{({\rm U}_+ - {\rm U}_-)^2}	\nonumber	\\
&	&		\nonumber	\\
& + & \frac{ f( {\rm U}_+ , {\rm U}_+ ) (1 - {\rm U}_-^2) \ U_-^a U_-^b
}{({\rm U}_+ - {\rm U}_-)^2}			\nonumber	\\
&	&		\nonumber	\\
& - & \frac{ f( {\rm U}_+ , {\rm U}_- ) (1 - {\rm U}_+^2)^{\frac{1}{2}}
(1 - {\rm U}_-^2)^{\frac{1}{2}} ( U_+^a U_-^b + U_-^a U_+^b ) }{({\rm U}_+
- {\rm U}_-)^2},		\nonumber	
\end{eqnarray}
where
\begin{equation}
f( {\rm U}_1 , {\rm U}_2 ) \ = \  \epsilon  {\rm U}_1 {\rm U}_2 +
p_\varphi  \ . \label{eq:fuu}
\end{equation}

Clearly, in order to cast the SEMT in the form (\ref{eq:emtsum}), the mixed
term must be absent and therefore the counter-rotating tangential velocities
must be related by
\begin{equation}
f( {\rm U}_+ , {\rm U}_- ) \ = \ 0 \ , \label{eq:liga}
\end{equation}
where we assume that $|{\rm U}_\pm| \neq 1$. Then, assuming a given choice for the counter-rotating velocities in agreement                                                                                                                                                                                                                                                                                                                                                                                                                                                                                                                                                                                                                                                                                                                                                                                                                                                                                             with the above relation, we can write the SEMT as (\ref{eq:emtsum}) with
\begin{equation}
S^{ab}_\pm =  \epsilon_\pm  \ U_\pm^a U_\pm^b \ ,
\end{equation}
so that we have two counter-rotating dust fluids with  energy densities given by
\begin{equation}
\epsilon_\pm  = \left[ \frac{ 1 - {\rm U}_\pm ^2 }{{\rm U}_\mp - {\rm U}_\pm}
\right] {\rm U}_\mp \epsilon. \label{eq:enercon} 
\end{equation}

Thus the SEMT $S^{ab}$ can
be written as the superposition of two counter-rotating dust streams
if, and only if, the constraint (\ref{eq:liga})  admits a solution such that ${\rm U}_+ \neq {\rm U}_-$. This result is completely equivalent to the
necessary and sufficient condition obtained in reference \cite{FMP}.

Similarly, we can write the current density in both cases as (\ref{eq:corsum}) with
\begin{equation}
{\rm j}^a_\pm  = \sigma _\pm U_\pm^a, 
\end{equation}
where $\sigma _\pm$ are the counter-rotating rest-charge densities of the fluids which are given by
\begin{subequations}
\begin{eqnarray}
& \sigma _{e \pm} &= \frac {\rm J^0}{V^0} \left[\frac { \sqrt{1-{\rm U}^2_\pm }} {{\rm U}\mp - {\rm U}\pm }\right] {\rm U}\mp, \label{eq:sige}  \\
&                 &    \nonumber  \\
& \sigma _{m \pm} & = \frac {\rm J^1}{W^1} \left[\frac { \sqrt{1-{\rm U}^2_\pm }} {{\rm U}\pm - {\rm U}\mp }\right], \label{eq:sigm}  
\end{eqnarray}
\end{subequations}
respectively.  Note that the counter-rotating energy densities $\epsilon_\pm$    and electrostatic (magnetostatic) charge  densities $\sigma _{e \pm}  (\sigma _{m \pm})$ are not uniquely defined by the above relations, also for definite
values of ${\rm U}_\pm$.

Another quantity related with the counter-rotating motion is the specific
angular momentum of a particle rotating at a radius $r$, defined as
$h_\pm = g_{\varphi\varphi} U_\pm^\varphi$. We can write
\begin{equation}
h_\pm \ = \ \frac{r e^{- \Phi} {\rm U}_\pm}{\sqrt{1 - {\rm
U}_\pm^2}} . \label{eq:moman}
\end{equation}
This quantity can be used to analyze the stability of the disks against
radial perturbations. The condition of stability,
\begin{equation}
\frac{d(h^2)}{dr} \ > \ 0 \ ,
\end{equation}
is an extension of Rayleigh criteria of stability of a fluid in rest in
a gravitational field \cite{FLU}.

We shall now analyze the possibility of a complete determination of the
vectors $U_\pm^a$. As we can see, the constraint (\ref{eq:liga}) does
not determine ${\rm U}_\pm$ uniquely, and so there is a freedom in
the choice of $U_\pm^a$. A possibility, commonly assumed, is to take
the two counter-rotating fluids as circulating along electro-geodesics \begin{equation}
\frac 12 \epsilon _\pm g_{ab,r}U^a_\pm U^b_\pm = - \sigma _\pm F_{ra} U^a_\pm.
\label{eq:geo}
\end{equation}
Let $\omega_\pm = U_\pm^1/U_\pm^0$ be the angular velocities of the particles.
Using (\ref{eq:vels}), (\ref{eq:enercon}) and (\ref{eq:sige}), in electrostatic case (\ref{eq:geo}) takes the form 
\begin{equation}
g_{11,r}\omega ^2+g_{00,r}=   - \frac {2{\rm j}^0 V_0^2} {\epsilon} \psi_{,r} ,
\end{equation}
so that
\begin{equation}
\omega_\pm \ = \ \pm \ \omega \qquad , \qquad \omega^2 \ = \ - \
\frac{g_{00,r}}{g_{11,r}} \  - \frac {2{\rm j}^0 V_0^2} {\epsilon g_{11,r}} \psi_{,r},
\end{equation}
and similarly, using (\ref{eq:sigm})  instead of (\ref{eq:sige}) in magnetostatic case we obtain 
\begin{equation}
g_{11,r}\omega ^2+g_{00,r}=   - \frac {2{\rm j}^1 V_0^2} {\epsilon} A_{,r} ,
\end{equation}
so that
\begin{equation}
\omega_\pm \ = \ \pm \ \omega \qquad , \qquad \omega^2 \ = \ - \
\frac{g_{00,r}}{g_{11,r}} \  - \frac {2{\rm j}^1 V_0^2} {\epsilon g_{11,r}} A_{,r}.
\end{equation}

Note that in both cases, the two geodesic fluids circulate with equal and opposite velocities.

In order to see if the geodesic velocities agree with (\ref{eq:liga}),
we need to compute $f( {\rm U}_+ , {\rm U}_- )$. In terms of $\omega_\pm$
we get
\begin{equation}
{\rm U}_\pm \ = \ \pm \ {\rm U} \ = \ \pm \left[ \frac{V^0}{W^1} \right]
\omega \ ,
  \end{equation}
and so, using the Einstein-Maxwell equations (\ref{eq:ele}) and (\ref{eq:mag}) and the expressions (\ref{eq:emt1}) - (\ref{eq:cormag}) for the SEMT and the 
current density we can show that $f( {\rm U}_+ , {\rm U}_- )$ vanishes in both cases, so that the constraint (\ref{eq:liga}) is equivalent to
\begin{equation}
{\rm U}^2 \ = \ \frac{p_\varphi}{\epsilon} \ ,   \label{eq:vel} 
\end{equation}
as is commonly assumed in the works concerning counter-rotating
disks. We now have two counter-rotating  charged dust streams with equal energy densities 
\begin{equation}
\epsilon_\pm \ = \ \frac{\epsilon - p_\varphi}{2},
\end{equation}
specific angular momenta
\begin{equation}
h_\pm = r e^{-\Phi} \sqrt { \frac {p_\varphi}{\epsilon - p_\varphi}},
\end{equation}
 charge densities  
\begin{subequations}
\begin{eqnarray}
& \sigma _{e \pm} & = -\frac12 e^{-\Phi}{\rm j}_0 \sqrt{1-\frac {p_\varphi}{\epsilon}},          \label{eq:sim} \\  
&  &     \nonumber    \\
& \sigma _{m \pm} & = \frac {1}{2r} e^{\Phi}{\rm j}_1 \sqrt{ \frac {\epsilon}{p_\varphi}-1 },    \label{eq:sie}  
\end{eqnarray}
\end{subequations}
and velocities given by (\ref{eq:vel}). As we can see, in this case we have a complete determination of all the
quantities involved in the CRM.

\section{Some Simple CRM Models}

\subsection{CRM for Reissner-Nordstr\"{o}m like disks}

The simplest electrostatic solution of the Einstein-
Maxwell equations is the well-known Reissner-Nordstr\"{o}m solution \cite{KSHM}, which can be written as (\ref{eq:met}) with
\begin{subequations}
\begin{eqnarray}
\Phi    &=& \frac 12 \ln \left [ \frac {x^2-1}{(x+a)^2} \right ], \\
        & & \nonumber   \\
\Lambda &=&\frac 12 \ln \left [ \frac {x^2-1}{x^2-y^2}  \right ], \\
        & & \nonumber   \\
\psi    &=& \frac { \sqrt 2 b}{x+a},
\end{eqnarray}\label{eq:reis}
\end{subequations}
where $a= m/k$, $b=e/k$, with $k^2=m^2-e^2$, so that $a^2=1+b^2$. Here $m$ and $e$ are the mass and the charge, respectively. $x$ and $y$ are the prolate spheroidal coordinates, related to the Weyl coordinates by 
\begin{subequations}
\begin{eqnarray}
2kx  & = \sqrt {r^2+(z+z_0+k)^2} + \sqrt {r^2+(z+z_0-k)^2}, \label{eq:cooprox} \\
     &  \nonumber       \\
2ky  & = \sqrt {r^2+(z+z_0+k)^2} - \sqrt {r^2+(z+z_0-k)^2}.\label{eq:cooproy}
\end{eqnarray}
\end{subequations}

Note that we have displaced the origin of the $z$ axis in $z_0$. This solution can be genera\-ted, in these coordinates,  using the well-known 
complex potencial formalism proposed by Ernst \cite{E1,E2} from   Schwarzschild solution \cite{Sch}. Indeed for $b=0$, it goes over into the Schwarzschild solution. Thus, $b$ is the parameter governing the electric field. One can obtain its  magnetostatic counterpart  cumputing the magnetostatic potencial via 
\begin{subequations}
\begin{eqnarray}
A_{,x} & = kf^{-1}(1-y^2) \psi _{,y}, \\
       & \nonumber  \\
A_{,y} & = -kf^{-1}(x^2-1) \psi _{,x}.
\end{eqnarray}\label{eq:potmag}
\end{subequations}
Thus, we find
\begin{equation}
A = \sqrt 2 kby,  
\end{equation}
again with  $a^2-b^2=1$. From the above expressions we can compute the physical quantities associated with disk. We obtain \begin{equation}
\tilde {\epsilon} =  \frac {4\bar{y}(\bar{x}^2-1)(a\bar{x}+\bar{y}^2)}{(\bar{x}+a)^2(\bar{x}^2-\bar{y}^2)^{ 3/2}}, 
\end{equation}

\begin{equation}
\tilde {p}_\varphi = \frac {4\bar{x} \bar{y}(1-\bar{y}^2)}{(\bar{x}+a)(\bar{x}^2-\bar{y}^2)^{ 3/2}},
\end{equation}

\begin{equation}
\tilde{\rm j}_t =  \frac {2 \sqrt 2 b \bar{y} (\bar{x}^2-1) }{(\bar{x}^2-\bar{y}^2)^{1/2}(\bar{x}+a)^3 },
\end{equation}

\begin{equation}
{\rm j}_\varphi = -  \frac {2 \sqrt 2 b \bar{x} (1-\bar{y}^2) }{(\bar{x}^2-\bar{y}^2)^{1/2}(\bar{x}+a) },
\end{equation}
 where $\tilde {\epsilon} = k \epsilon$, $\tilde {p}_\varphi
= k p_\varphi$ and $\tilde{\rm j}_t = k {\rm j}_t$.
$\bar{x}$ and  $\bar{y}$ are given by
\begin{subequations}
\begin{eqnarray}
2\bar{x}  & = \sqrt {\tilde{r}^2+(\alpha +1)^2} + \sqrt {\tilde{r}^2+(\alpha-1)^2}, \label{eq:xbar} \\
     &  \nonumber       \\
2\bar{y}  & = \sqrt {\tilde{r}^2+(\alpha+1)^2} - \sqrt {\tilde{r}^2+(\alpha-1)^2},\label{eq:ybar}
\end{eqnarray}
\end{subequations}
where $\tilde{r}=r/k$ and $\alpha = z_0/k$, with $\alpha >1$.

\begin{figure}
$$
\begin{array}{cc}
\tilde {\epsilon}  & \tilde {p}_\varphi \\
\epsfig{width=3in,file=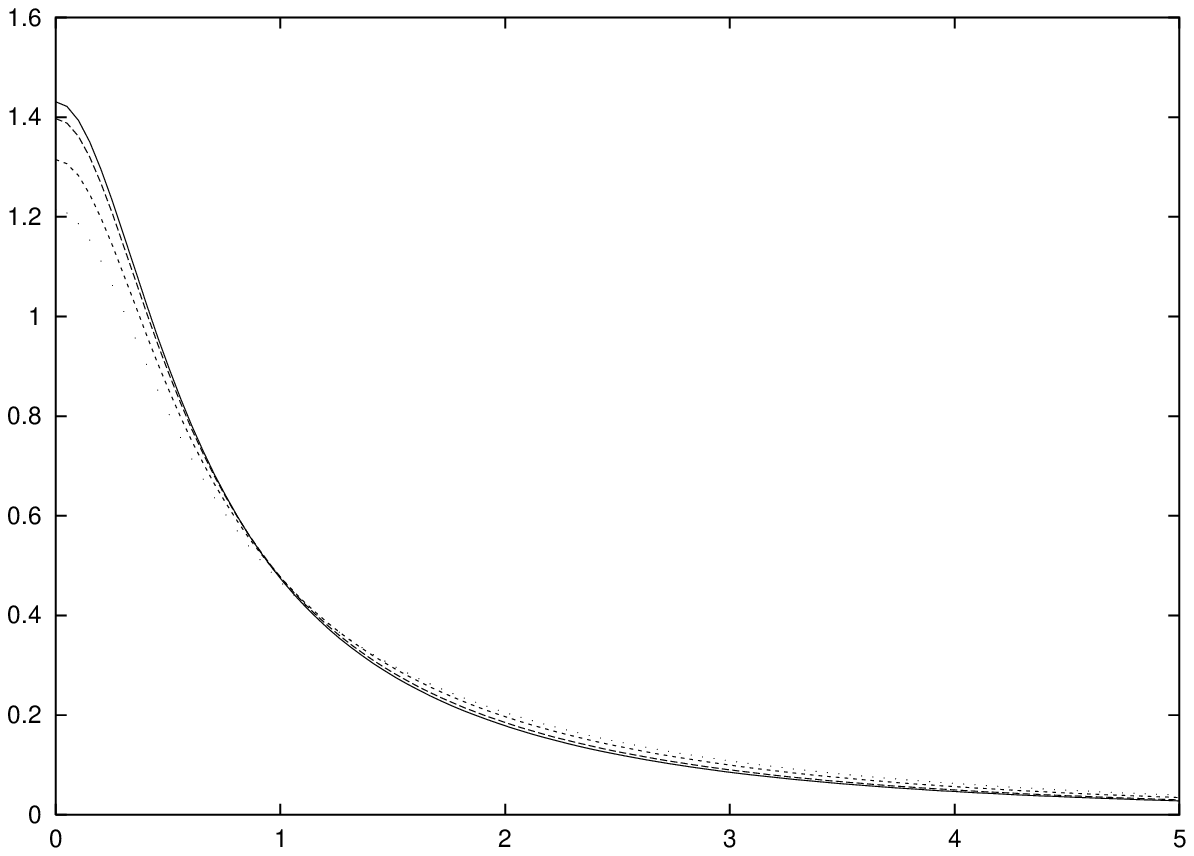} & \epsfig{width=3in,file=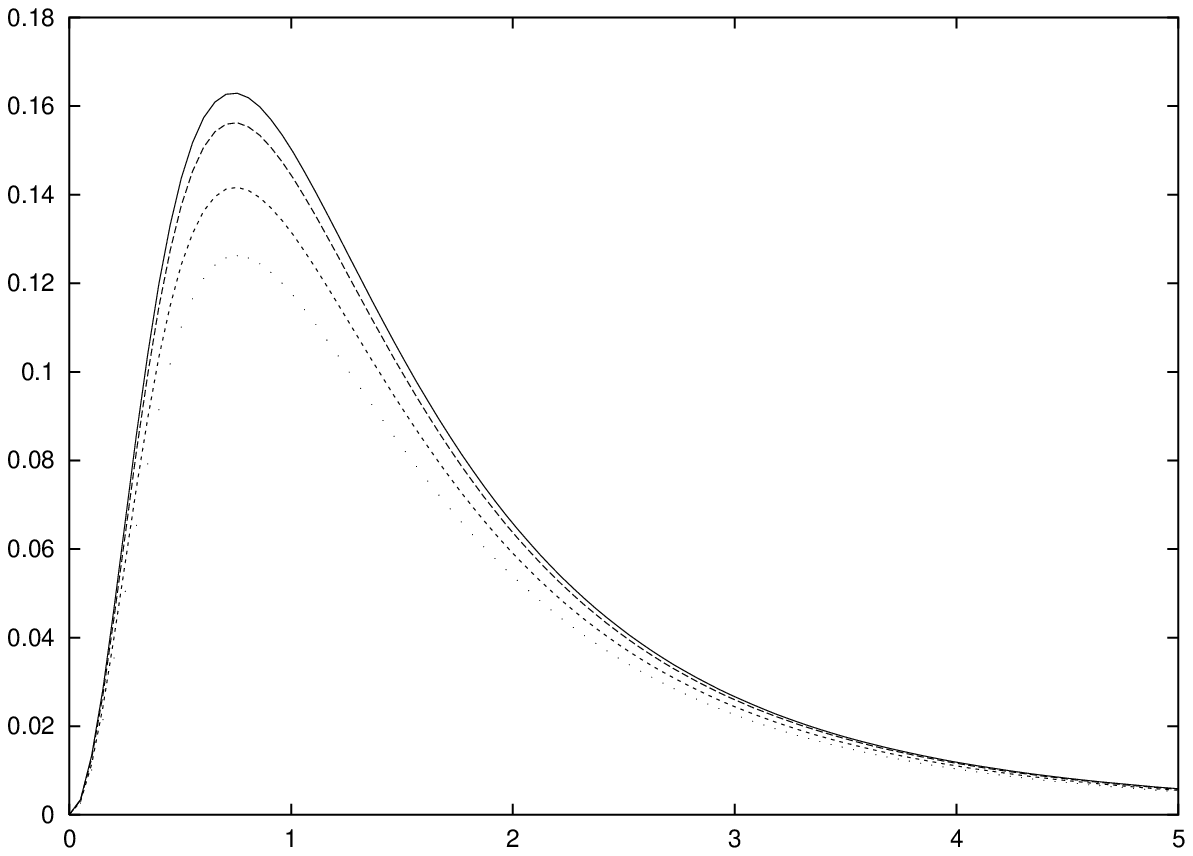} \\
\tilde r & \tilde r    \\
(a)     &   (b)
\end{array}
$$	
\caption{$(a)$ Energy density $\tilde {\epsilon}$ and  $(b)$ azimuthal pressure  $\tilde {p}_\varphi$ as functions of $\tilde{r}$ for disks  with $\alpha=1.5$ and $b=0$ (solid curve), $0.5$, $1.0$, and $1.5$ (dotted curve).}\label{fig:renerpres}
\end{figure}

In order to study the behavior of these quantities we perform a graphical analisis of them for disks  with $\alpha=1.4$ and $b=0$, $0.5$, $1.0$, and $1.5$.  In Fig. \ref{fig:renerpres}$(a)$
we show  the energy density  $\tilde {\epsilon}$ as a function of $\tilde{r}$. We  see that the energy density present a maximum at $\tilde {r} = 0$ and then decreases rapidly with  $\tilde{r}$. We also see that  the presence of electric (magnetic) field decreases the energy density  in the central regions of the disks and  later increases it. For other values of $\alpha$ and $b$ we obtain  a similar behavior.

In Fig. \ref{fig:renerpres}$(b)$ we plot  the azimuthal pressure  $\tilde {p}_\varphi$ as function of $\tilde{r}$. We can observe that the pressure increases rapidly as one moves away from the disk center, reaches a maximum and  later rapidly decreases.  Note  that electric (magnetic) field decreases the pressure everywhere on the disk.  For other values of $\alpha$ and $b$ we have  a similar behavior.

\begin{figure}
$$
\begin{array}{cc}
\tilde {\rm j}_t   &  \rm j_\varphi   \\
\epsfig{width=3in,file=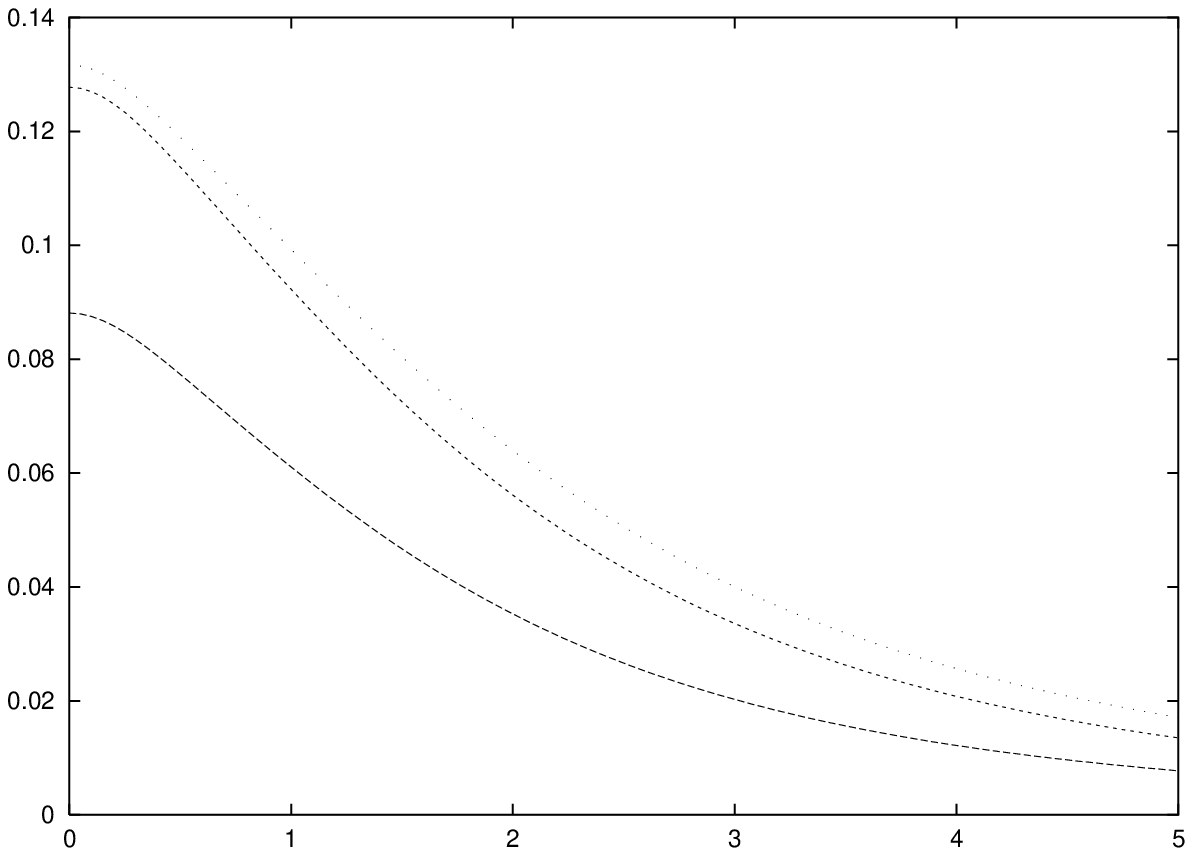} &  \epsfig{width=3in,file=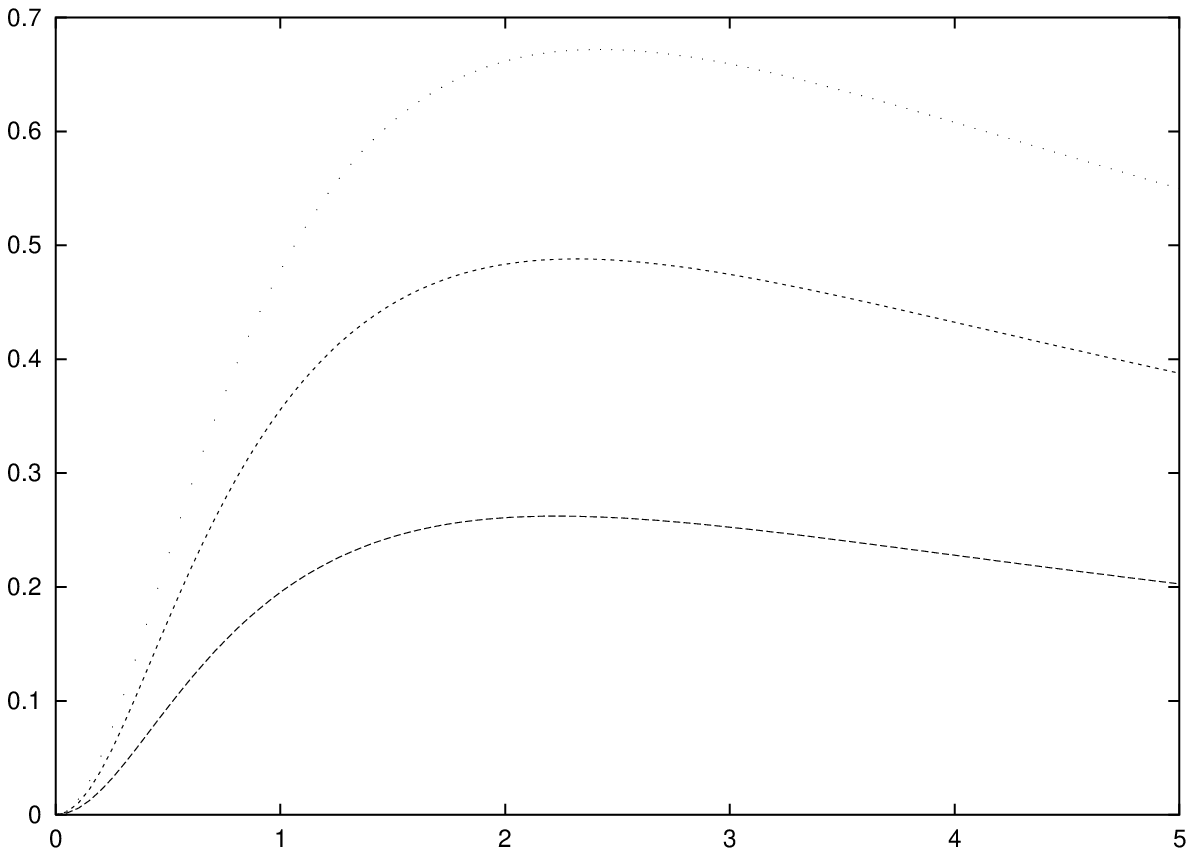} \\
\tilde r  &  \tilde r  \\
(a)   &   (b)
\end{array}
$$	
\caption{ Surface current density:  $(a)$ $\tilde {\rm j}_t$ and $(b)$   ${\rm j}_\varphi$ as functions of $\tilde r$ for disks  with $\alpha=1.5$ and $b=0$ (axis  $\tilde r$), $0.5$, $1.0$, and $1.5$ (dotted curve).}\label{fig:rcorelecmag}
\end{figure}

The electrostatic current $\tilde {\rm j}_t$ is presented in Fig. \ref{fig:rcorelecmag}$(a)$.    Similarly to the energy density, it has a maximum at the disk center and then decreases rapidly with  $\tilde{r}$. For other values of $\alpha$ and $b$ we observe  a similar behavior. In addition, the magnetostatic current ${\rm j}_\varphi$ is ploted  in Fig. \ref{fig:rcorelecmag}$(b)$  as function also of $\tilde{r}$. As we can observe it has  a similar behavior to the pressure.

\begin{figure}
$$
\begin{array}{cc}
\rm U^2  & \rm U^2   \\
\epsfig{width=3in,file=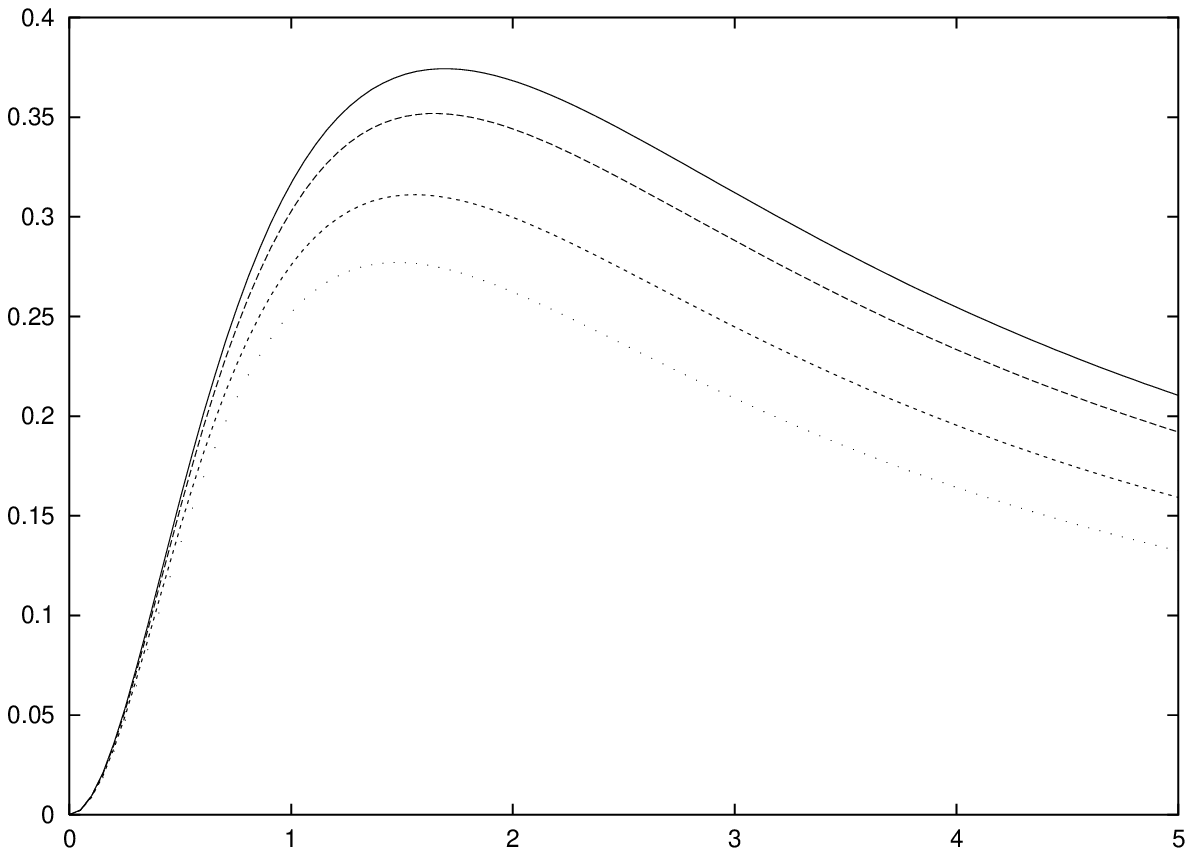} &  \epsfig{width=3in,file=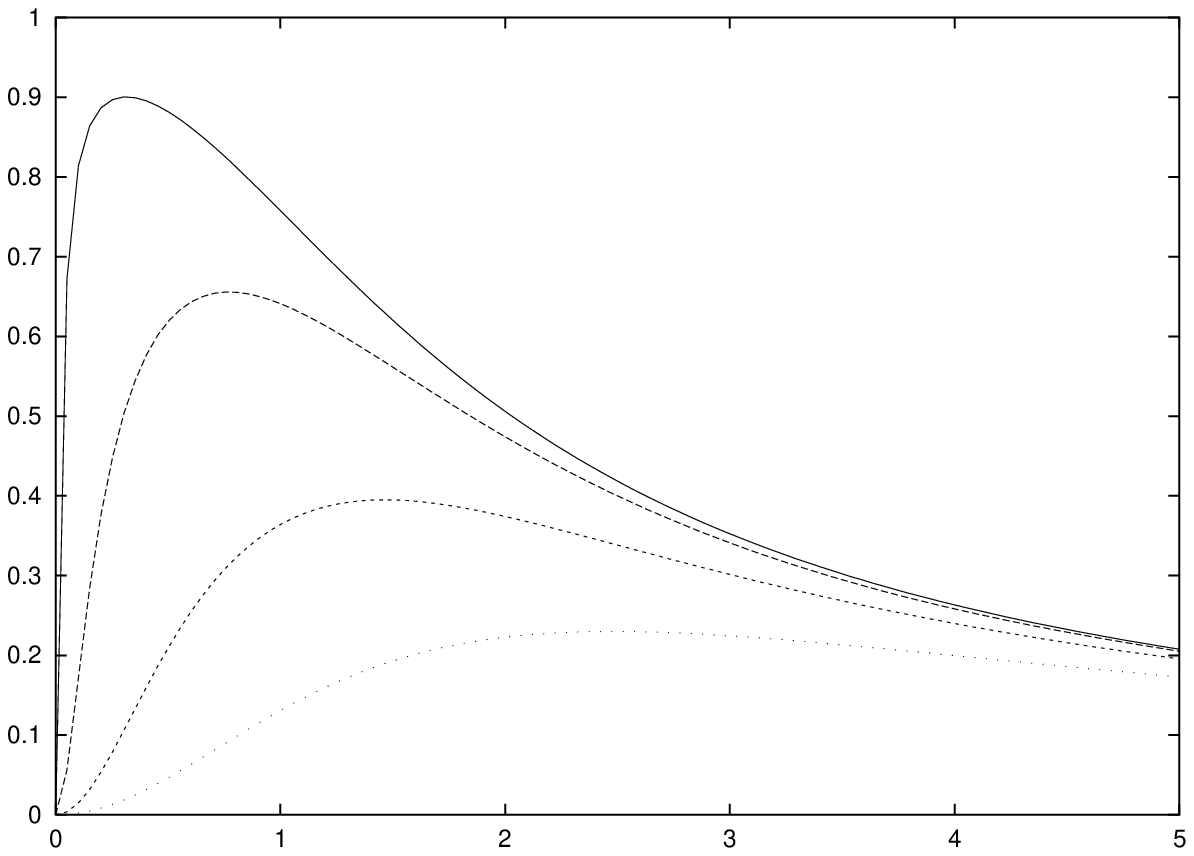} \\
\tilde r  & \tilde r \\
(a)  & (b)
\end{array}
$$	
\caption{ Tangential velocity $\rm U ^2$   as function of $\tilde r$ for disks  with $(a)$ $\alpha=1.5$ and $b=0$ (solid curve), $0.5$, $1.0$, and $1.5$ (dotted curve), and $(b)$  $b = 0.5$ and $\alpha = 1.01$ (solid curve), $1.1$, $1.4$, $2.0$ (dotted curve)} \label{fig:rvel}
\end{figure}

We now consider the  CRM for the same value of the parameters. All the significant quantities can also be expresed in analytic form from above expressions but the results are so cumbersome that it is best just to analyze them graphically.
In Fig.  \ref{fig:rvel}$(a)$ we show the velocity curves of counter-rotating streams ${\rm U}^2$ as functions of $\tilde{r}$. We observe that it increases rapidly in the central region of the disk, achieves a maximum and  later  decreases monotonly.  We also see that  the presence  of electric (magnetic) field makes  least  relativistic the disks. In addition, in Fig. \ref{fig:rvel}$(b)$, we plot ${\rm U}^2$ for disks with $b = 0.5$ and $\alpha = 1.01$, $1.1$, $1.4$, and $2.0$. We find that the disks become least relativistic with increasing $\alpha$. We also find that the disks with   $\alpha <1$  cannot be built from  CRM  because $\rm{U}^2>1$  (not shown in the figure).

\begin{figure}
$$
\begin{array}{cc}
\tilde h^2 &  \tilde \epsilon _\pm \\
\epsfig{width=3in,file=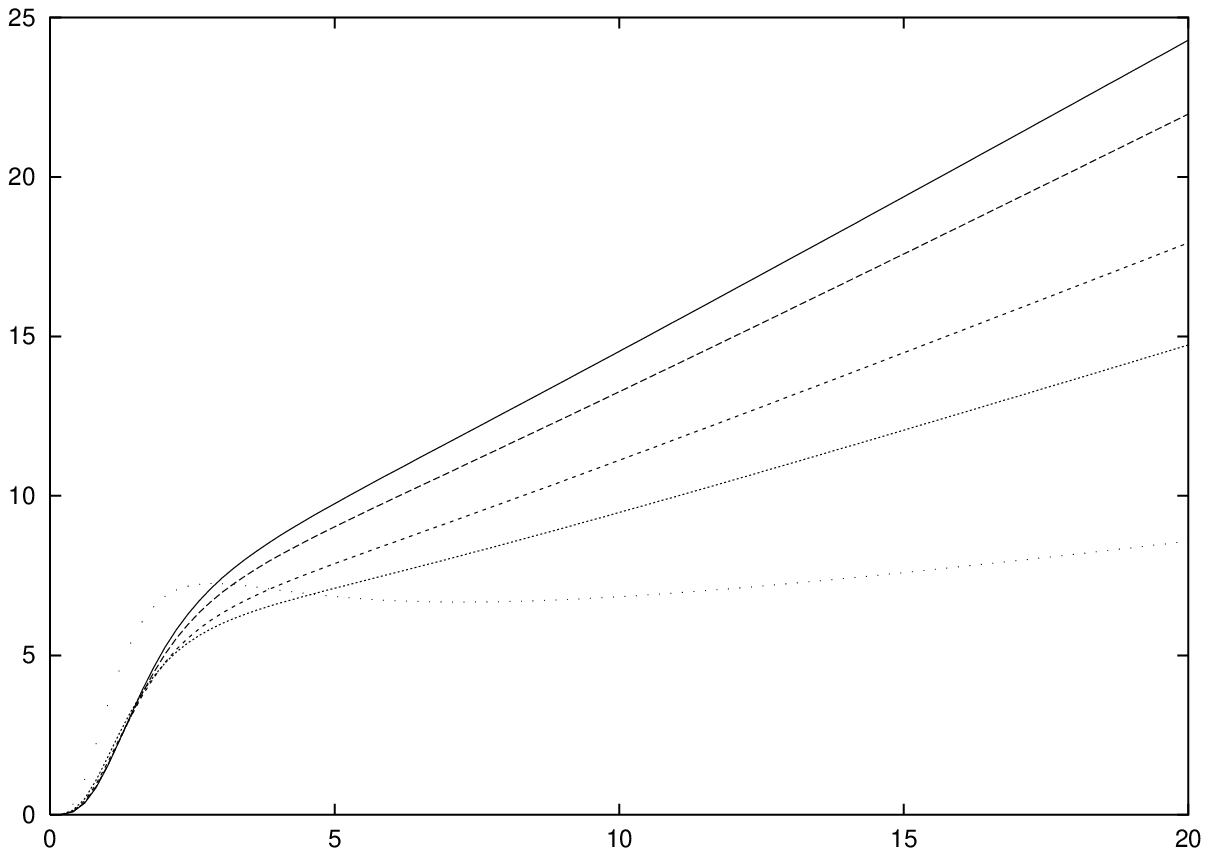} & \epsfig{width=3in,file=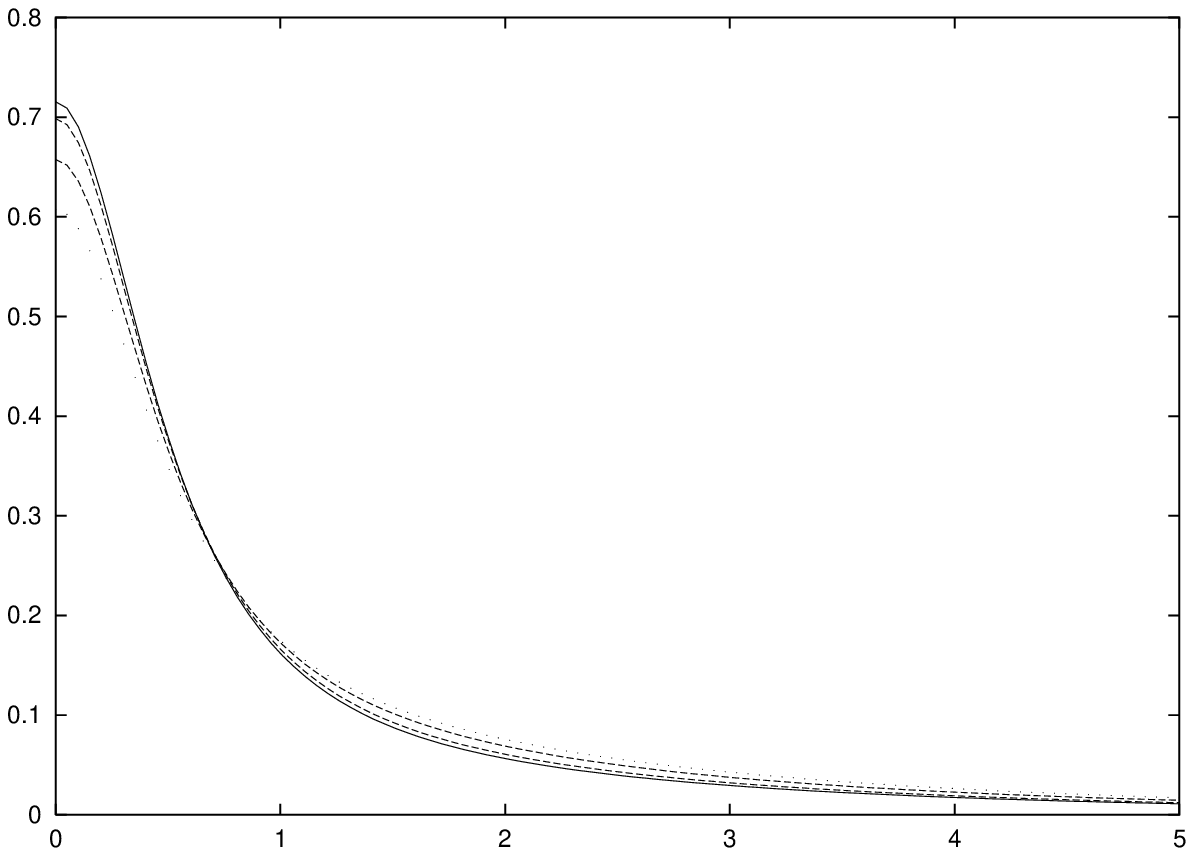}  \\
\tilde r & \tilde r \\
(a)  &  (b)
\end{array}
$$	
\caption{$(a)$  Specific angular momentum $\tilde h^2$ as function of $\tilde r$ for disks  with $\alpha=1.5$ and $b=0$ (solid curve), $0.5$, $1.0$, $1.5$ and $4.0$ (dotted curve). $(b)$ Mass densities $\tilde \epsilon _\pm$ as function of $\tilde r$ for disks  with $\alpha=1.5$ and $b=0$ (solid curve), $0.5$, $1.0$, and $1.5$ (dotted curve).}\label{fig:rmoencon}
\end{figure}

\begin{figure}
$$
\begin{array}{cc}
- \tilde \sigma _{e \pm} &  \tilde \sigma _{m \pm}  \\
\epsfig{width=3in,file=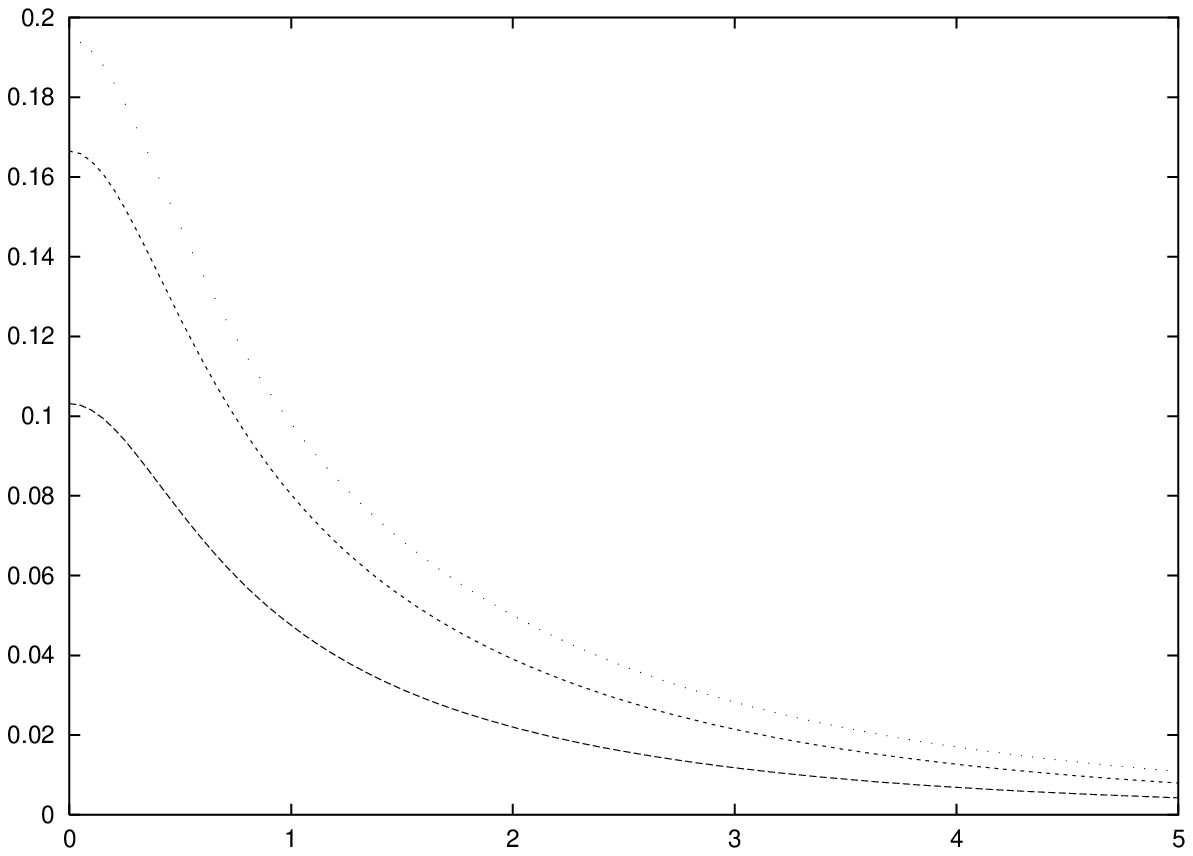} & \epsfig{width=3in,file=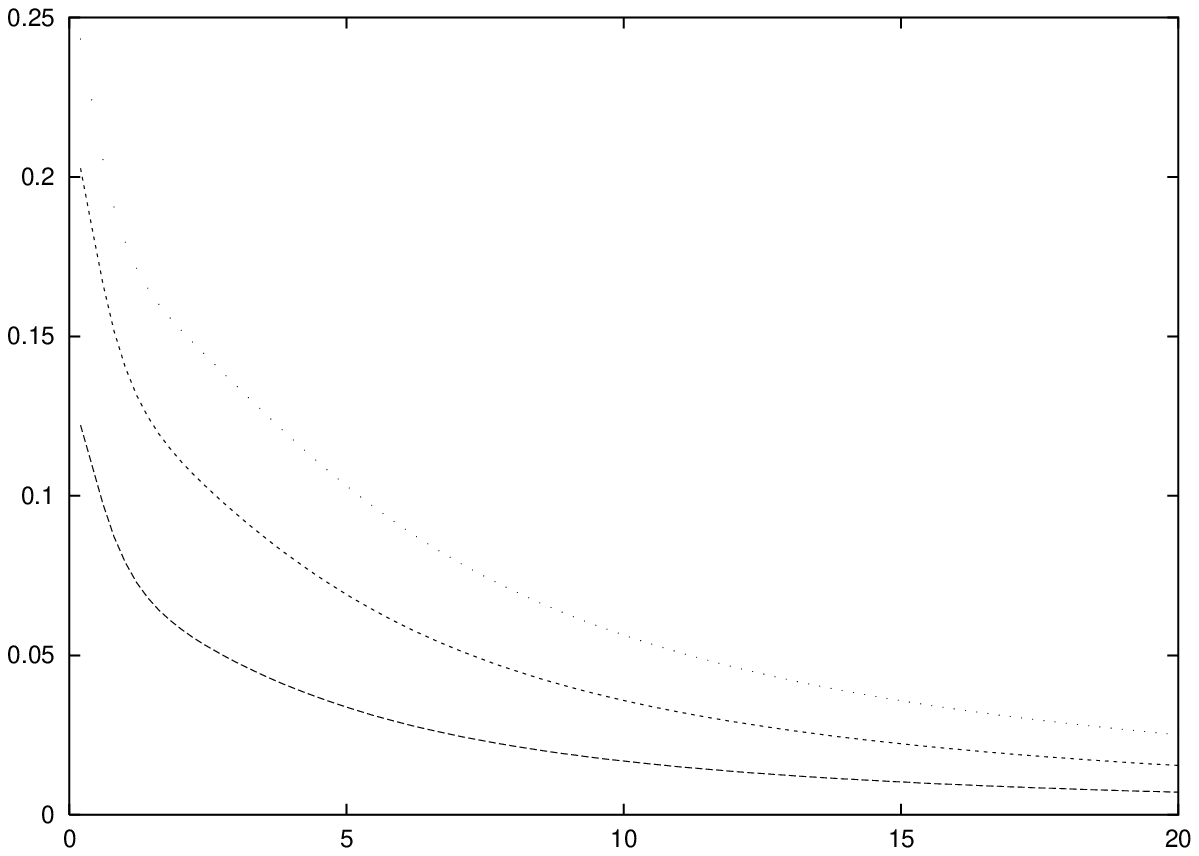}  \\
\tilde r  & \tilde r \\
(a)  & (b)
\end{array}
$$	
\caption{Charge densities:  $(a)$ $\tilde \sigma _{e \pm}$ and  $(b)$   $\tilde \sigma _{m \pm}$ as functions of $\tilde r$ for disks  with $\alpha=1.5$ and $b=0$ (axis $\tilde r$), $0.5$, $1.0$, and $1.5$ (dotted curve).}\label{fig:rdenelecmag}
\end{figure}

In Fig. \ref{fig:rmoencon}$(a)$  we have drawn the specific angular momentum $\tilde h^2$ of counter-rotating fluids, where $\tilde h =  h_\pm /k$. In the cases considered we obtain $\tilde h^2$ as increasing monotonic functions of $\tilde{r}$ what correspond to stable CRM for the disks. However, the CRM cannot be applied for $b=4.0$ (dotted curve). Thus  the presence of electric (magnetic) field  can makes unstable the CRM against radial perturbations.
Finally, in Figs. \ref{fig:rmoencon}$(b)$ and \ref{fig:rdenelecmag}  the plots of the  mass densities  $\epsilon _\pm$ and  electrostatic (magnetostatic) charge  densities $\sigma _{e \pm}  (\sigma _{m \pm})$ of both streams are shown. These present a maximun at the disks center and then decrease monotonly. Therefore,  the CRM constructed from this value of the parameters are well behaved.

\subsection{CRM for Taub-NUT like disks}

A Taub-NUT like solution to the Einstein-Maxwell equations is  
\begin{subequations}
\begin{eqnarray}
\Phi    &=& \ln \left [ \frac {x^2-1}{x^2+2ax+1} \right ], \\
        & & \nonumber   \\
\Lambda &=& 2 \ln \left [ \frac {x^2-1}{x^2-y^2}  \right ], \\
        & & \nonumber   \\
\psi    &=& \frac {2 \sqrt 2 bx}{x^2+2ax+1}, 
\end{eqnarray}\label{eq:reis}
\end{subequations}
with  $a^2-b^2=1$. $x$ and $y$ are, again, the prolate spheroidal coordinates, given by Eqs. (\ref{eq:cooprox}) and (\ref{eq:cooproy}). This solution can also be generated, in these coordinates,  using the well-known 
complex potencial formalism proposed by Ernst \cite{E1,E2} from    Weyl 2-solution (Darmois \cite{KSHM}). Indeed for $b=0$, it goes over into the Darmois solution . Thus, again,  $b$ is the  electric parameter. One  can also obtain its  magnetostatic counterpart  cumputing the magnetostatic potencial using (\ref{eq:potmag}), and we get
\begin{equation}
A = 2 \sqrt 2 kby,  
\end{equation}
again with  $a^2-b^2=1$. The physical quantities associated with the disk now can be written as 
\begin{equation}
\tilde {\epsilon} = \frac {8\bar{y}[\bar{x}(a\bar{x}^3-3a\bar{x}
-2)+\bar{y}^2(2\bar{x}^3+3a\bar{x}^2-a)]}{(\bar{x}^2-1)(\bar{x}^2+2a\bar{x}+1)^2},     
\end{equation}

\begin{equation}
\tilde {p}_\varphi = \frac {16\bar{x}\bar{y}(1-\bar{y}^2)}{(\bar{x}^2-1)(\bar{x}^2+2a\bar{x}+1)},
\end{equation}

\begin{equation}
\tilde{\rm j}_t = \frac {4  \sqrt 2 b\bar{y}(\bar{x}^2-1)(\bar{x}^2-\bar{y}^2)}{(\bar{x}^2+2a\bar{x}+1)^3},   
\end{equation}

\begin{equation}
{\rm j}_\varphi=-\frac{4  \sqrt 2 b\bar{x}(1-\bar{y}^2)(\bar{x}^2-\bar{y}^2)}{(\bar{x}^2-1)(\bar{x}^2+2a\bar{x}+1)},
\end{equation}
where $\bar{x}$ and  $\bar{y}$ are given by Eqs. (\ref{eq:xbar}) and (\ref{eq:ybar}).

\begin{figure}
$$
\begin{array}{cc}
\tilde {\epsilon} &   \tilde {p}_\varphi \\
\epsfig{width=3in,file=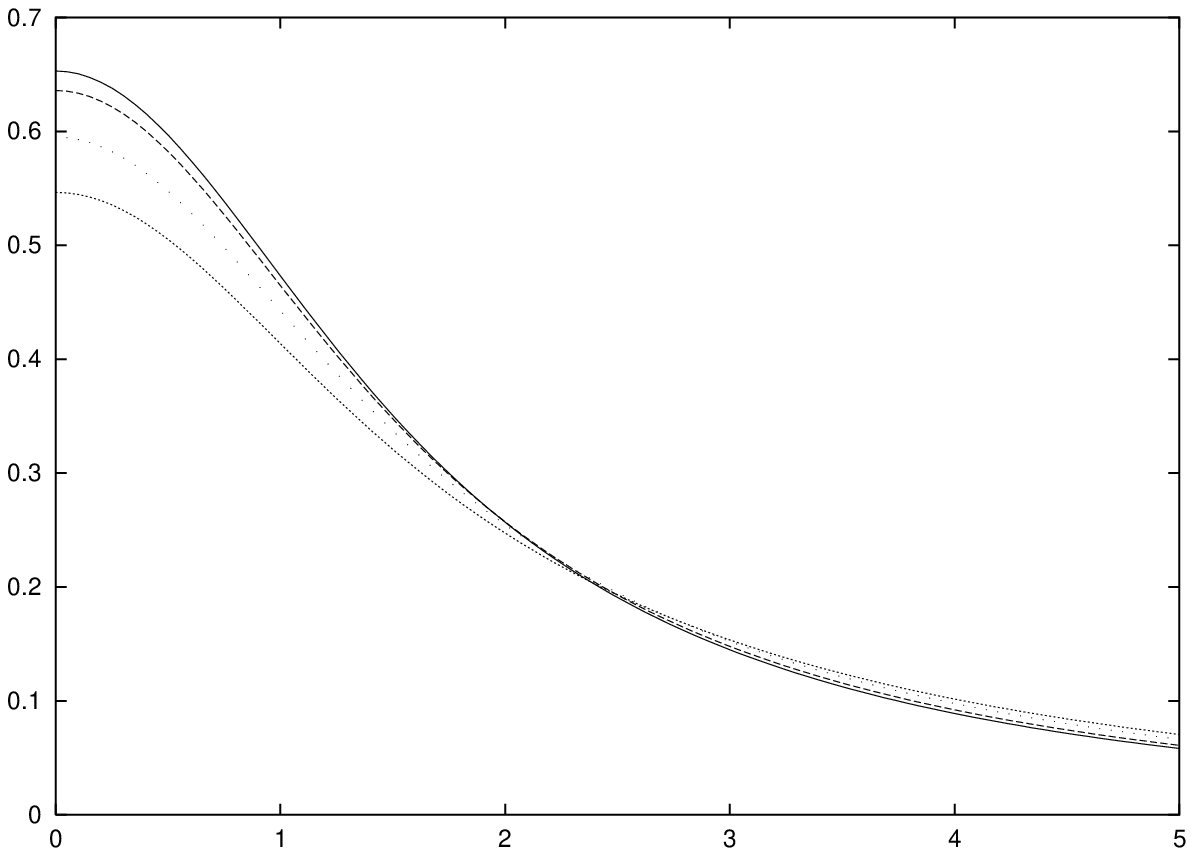} &  \epsfig{width=3in,file=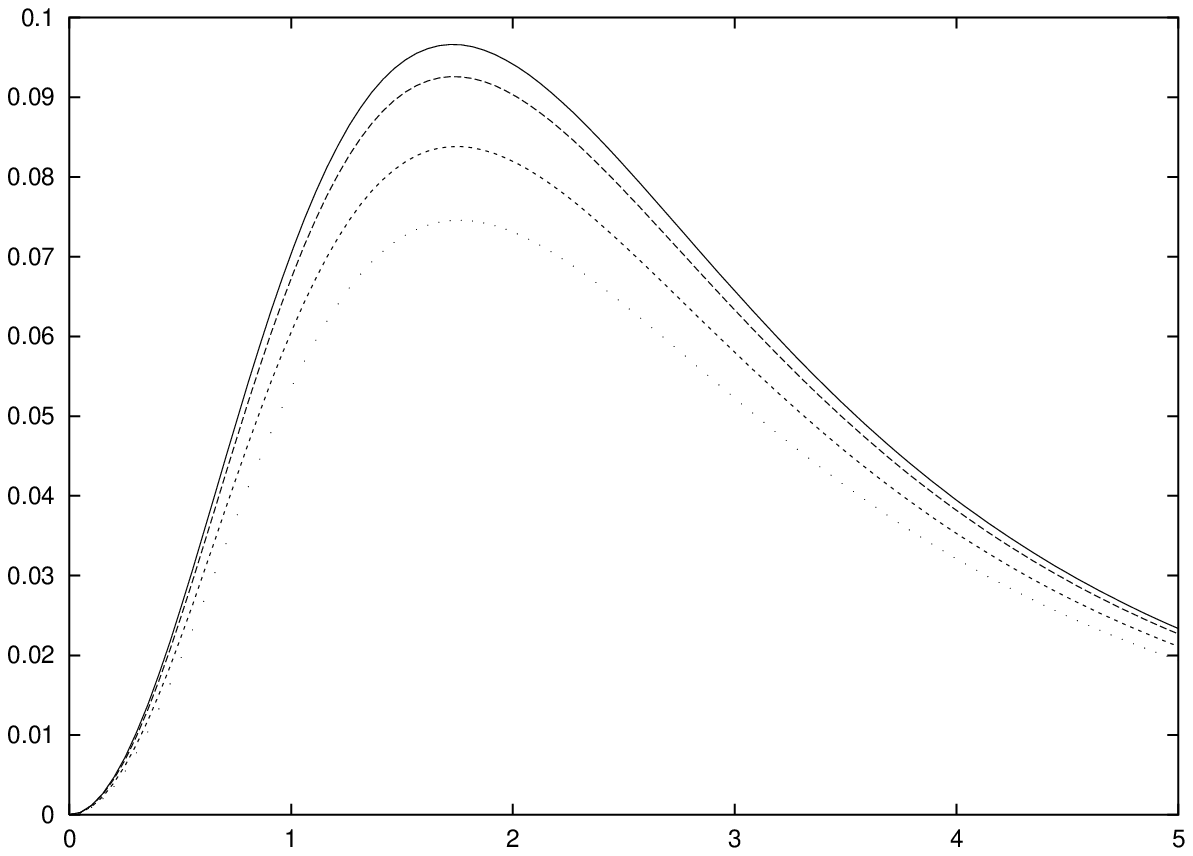} \\
\tilde r & \tilde r \\
(a)   &  (b)
\end{array}
$$	
\caption{$(a)$ Energy density  $\tilde {\epsilon}$ and $(b)$ azimuthal pressure  $\tilde {p}_\varphi$ as functions of $\tilde{r}$ for disks  with $\alpha=2.5$ and $b=0$ (solid curve), $0.5$, $1.0$, and $1.5$ (dotted curve).}\label{fig:tenerpres}
\end{figure}

\begin{figure}
$$
\begin{array}{cc}
\tilde {\rm j}_t & \rm j_\varphi \\
\epsfig{width=3in,file=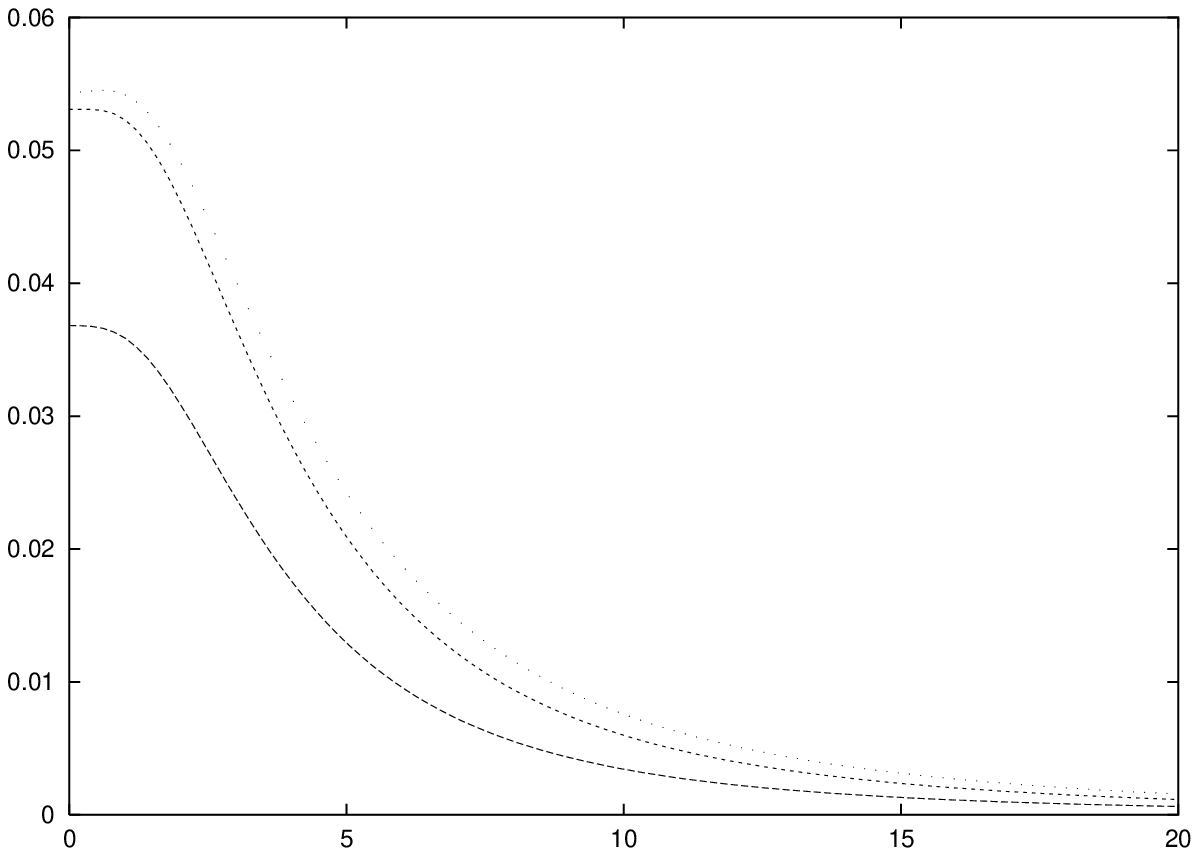} & \epsfig{width=3in,file=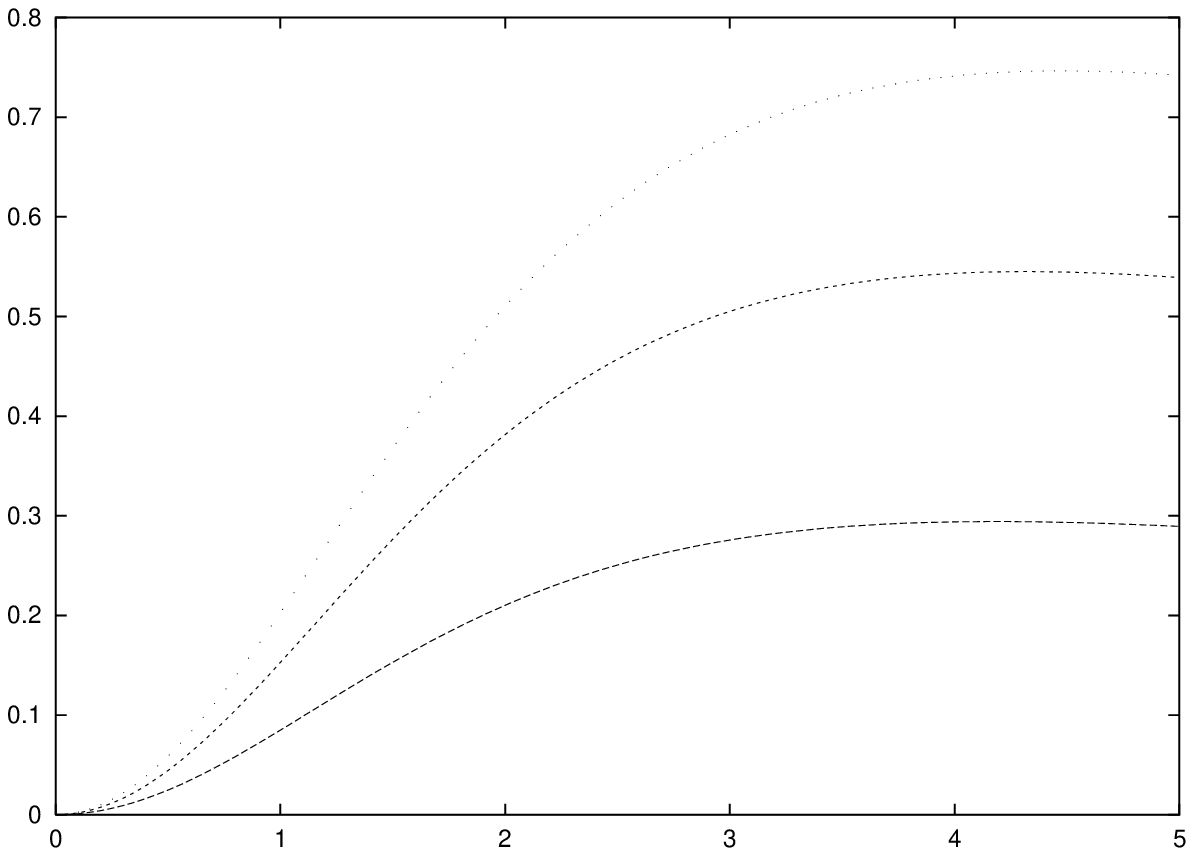} \\
\tilde r  & \tilde r \\
(a)  & (b)
\end{array}
$$	
\caption{Planar current density: $(a)$  $\tilde {\rm j}_t$ and $(b)$  $ {\rm j}_\varphi$ as functions of $\tilde r$ for disks  with $\alpha=2.5$ and $b=0$ (axis  $\tilde r$), $0.5$, $1.0$, and $1.5$ (dotted curve).}\label{fig:tcorelecmag}
\end{figure}

\begin{figure}
$$
\begin{array}{cc}
\rm U^2  & \rm U^2   \\
\epsfig{width=3in,file=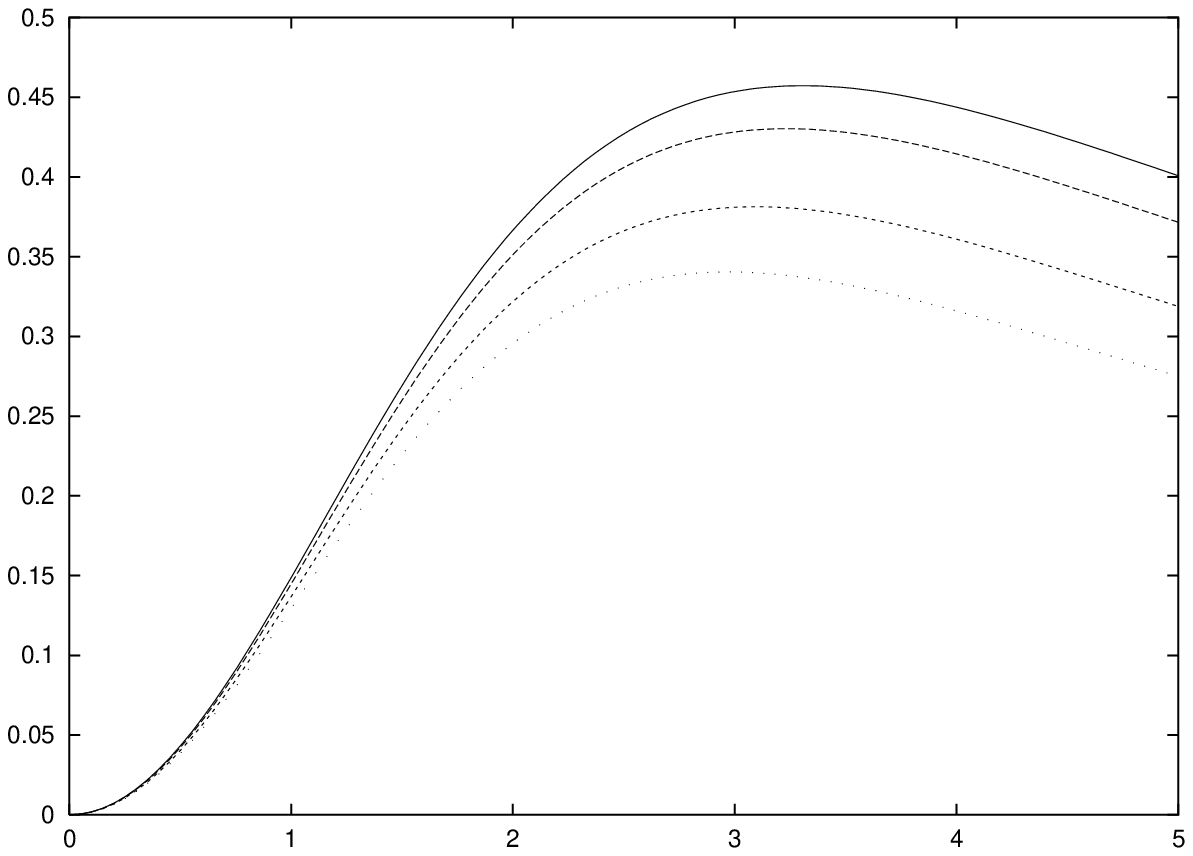} &  \epsfig{width=3in,file=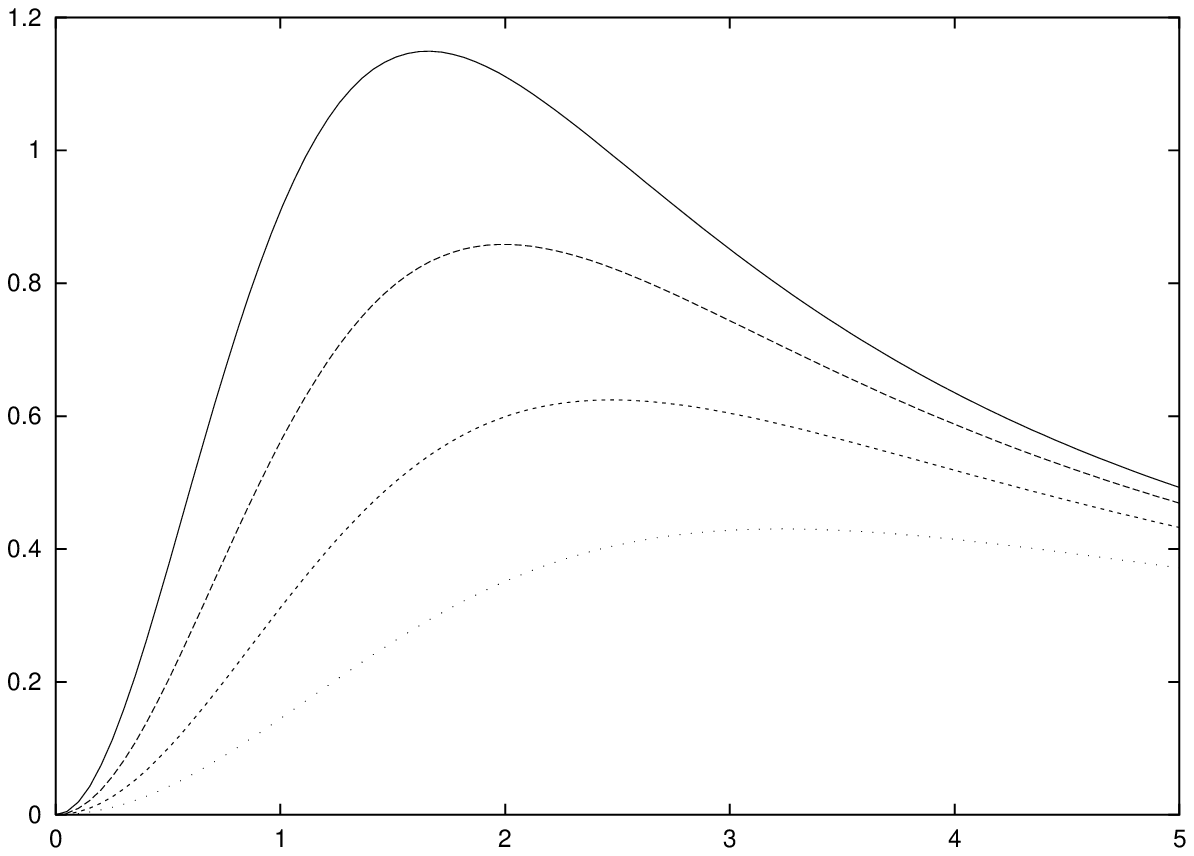} \\
\tilde r  & \tilde r \\
(a)  & (b)
\end{array}
$$	
\caption{ Tangential velocity $\rm U ^2$   as function of $\tilde r$ for disks  with $(a)$ $\alpha=2.5$ and $b=0$ (solid curve), $0.5$, $1.0$, and $1.5$ (dotted curve), and $(b)$  $b = 0.5$ and $\alpha = 1.5$ (solid curve), $1.7$, $2.0$, $2.5$ (dotted curve)} \label{fig:tvel}
\end{figure}

\begin{figure}
$$
\begin{array}{cc}
\tilde h^2 &  \tilde \epsilon _\pm   \\
\epsfig{width=3in,file=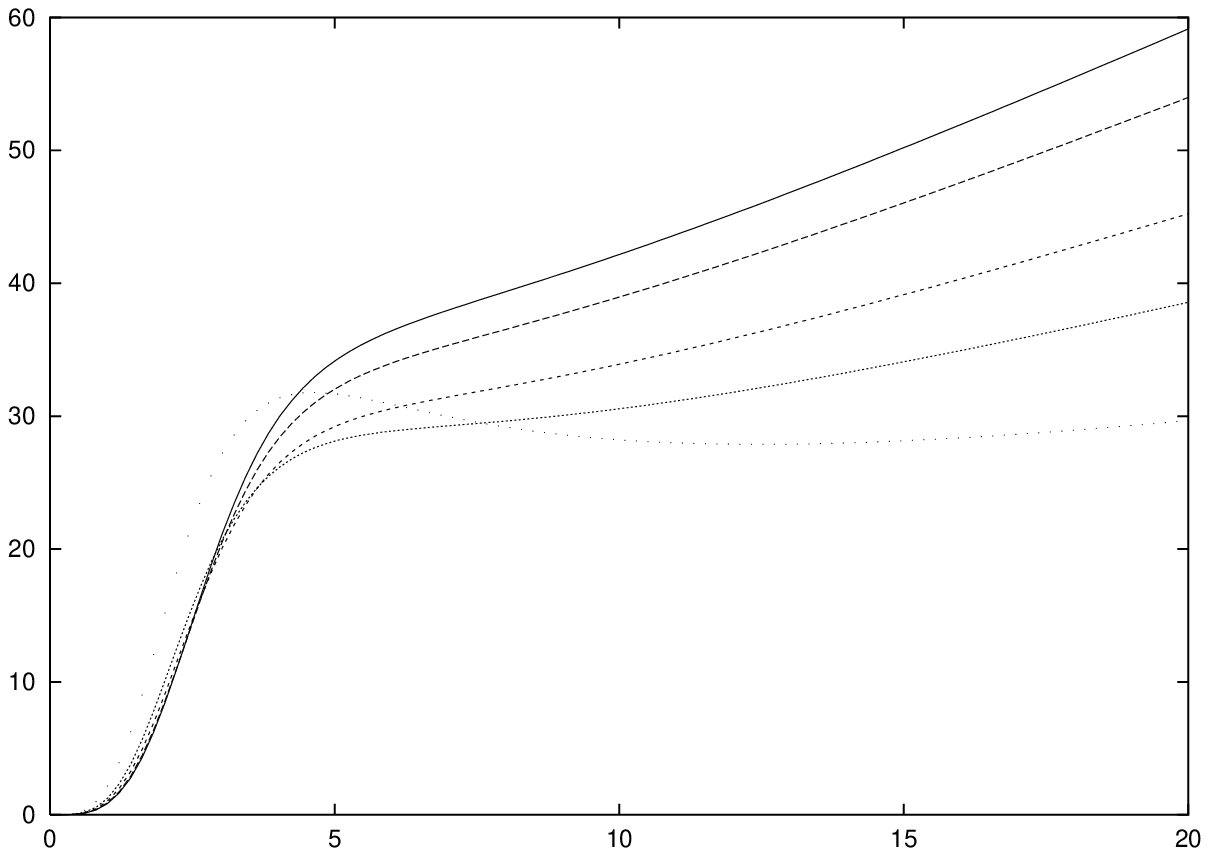} & \epsfig{width=3in,file=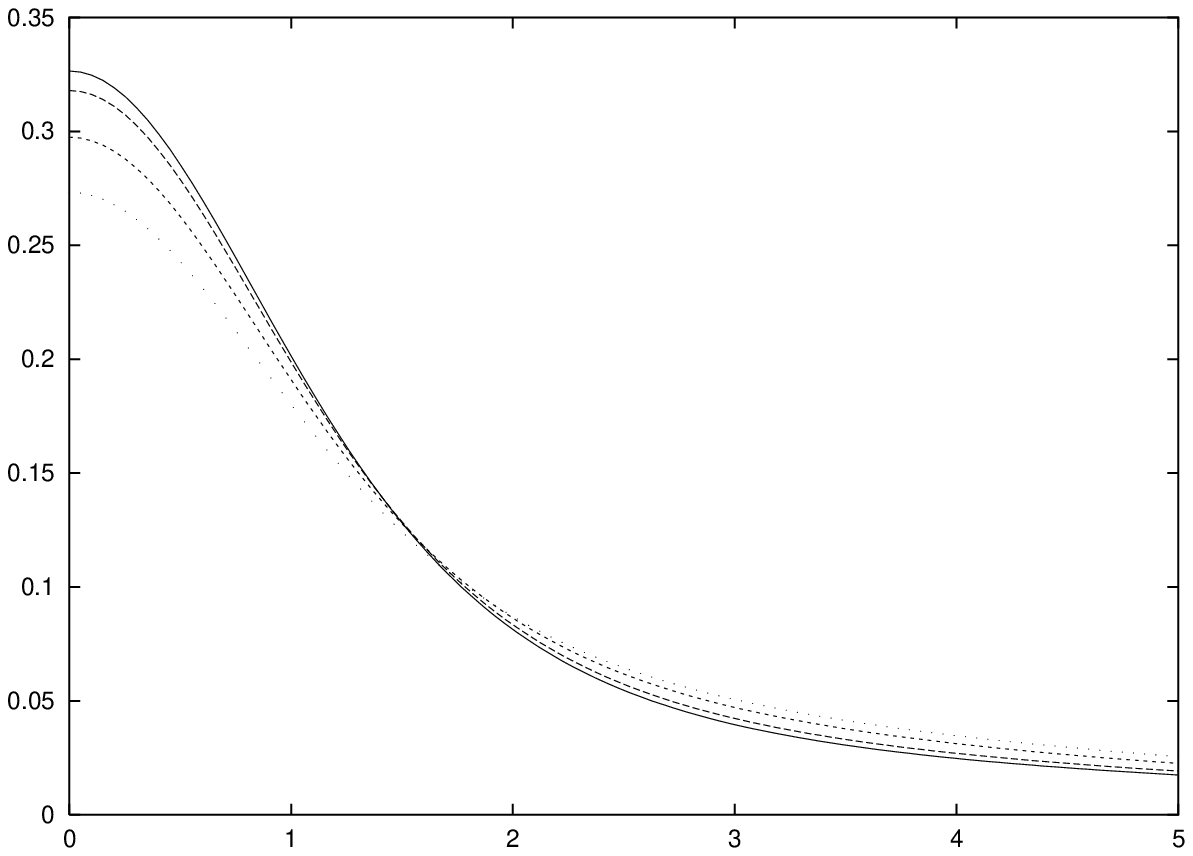} \\
\tilde r & \tilde r \\
(a) & (b)       
\end{array}
$$	
\caption{$(a)$ Specific angular momentum $\tilde h^2$  as function of $\tilde r$ for disks  with $\alpha=2.5$ and $b=0$ (solid curve), $0.5$, $1.0$, $1.5$, and $3.0$ (dotted curve).  $(b)$ Mass densities  $\tilde \epsilon _\pm$ as function of $\tilde r$ for disks  with $\alpha=2.5$ and $b=0$ (solid curve), $0.5$, $1.0$, and $1.5$ (dotted curve). }\label{fig:tmoencon}
\end{figure}

\begin{figure}
$$
\begin{array}{cc}
- \tilde \sigma _{e \pm} & \tilde \sigma _{m \pm} \\
\epsfig{width=3in,file=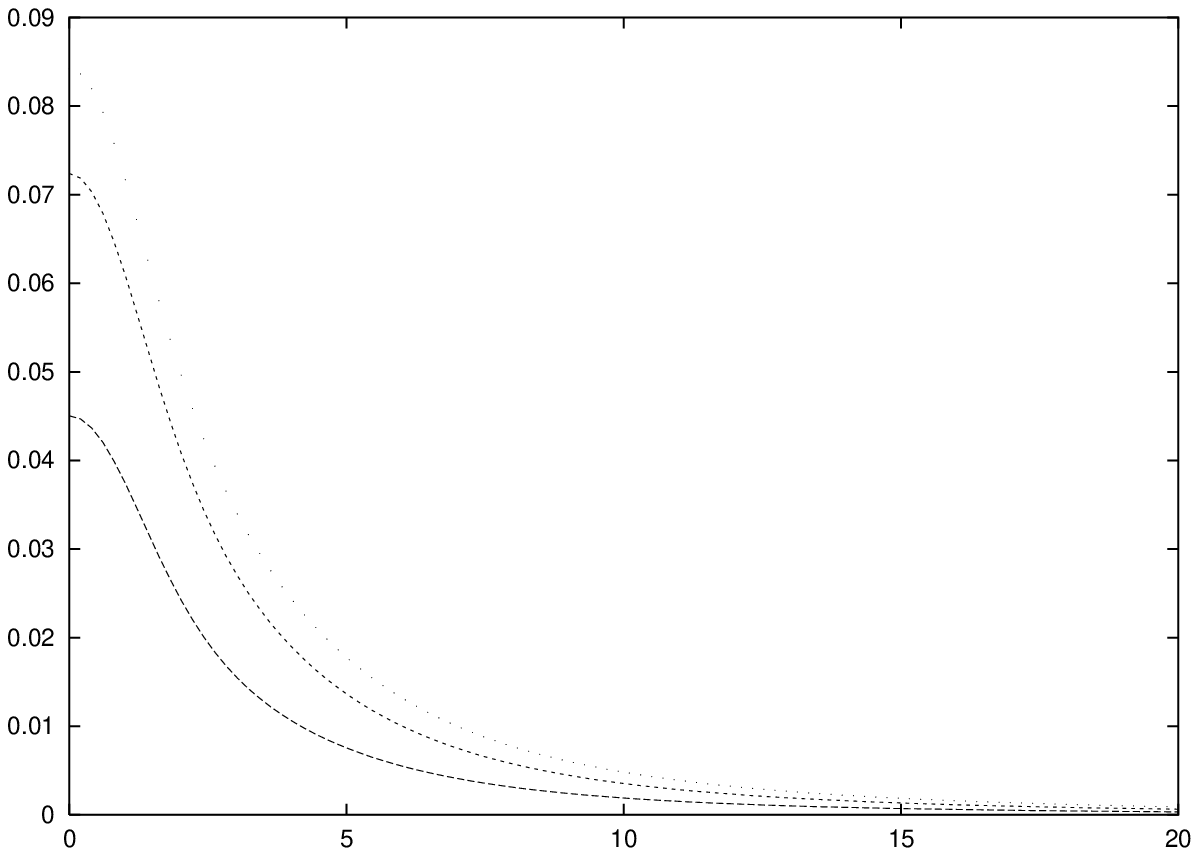} & \epsfig{width=3in,file=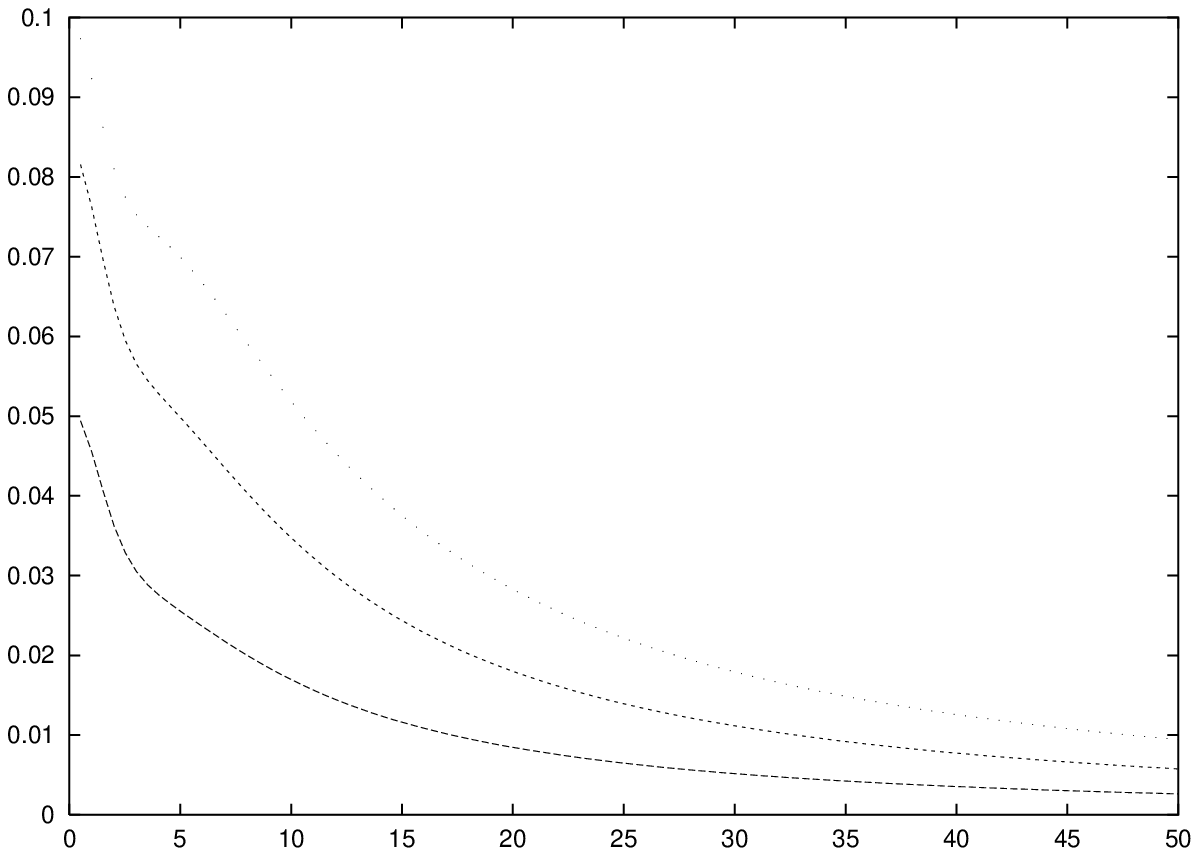} \\
\tilde r & \tilde r \\
(a)   & (b)
\end{array}
$$	
\caption{Charge densities: $(a)$ $\tilde \sigma _{e \pm}$ and $(b)$ $\tilde \sigma _{m \pm}$ as functions of $\tilde r$ for disks  with $\alpha=2.5$ and $b=0$ (axis $\tilde r$), $0.5$, $1.0$, and $1.5$ (dotted curve).}\label{fig:tdenelecmag}
\end{figure}

In Figs.  \ref{fig:tenerpres} and \ref{fig:tcorelecmag} the plots of the quantities  $\tilde {\epsilon}$, $\tilde {p}_\varphi$, $\tilde{\rm j}_t$ and ${\rm j}_\varphi$ are presented for disks with  $\alpha=2.5$ and $b=0$, $0.5$, $1.0$, and $1.5$. We see that these disks have a similar  behavior to the previous case. Equally, the  relevant  quantities of the CRM are shown in following figures  for the same value of the parameters. Theses CRM are more relativistic than the ones built from  Reissner-Nordstr\"{o}m like solution (Fig. \ref{fig:tvel}(a) ). Note  that (Fig. \ref{fig:tvel}(b) ) the disks with $b=0.5$ $\alpha = 1.5 $ (solid curve)  cannot be constructed from the CRM because $\rm{U}^2>1$.
We also note that  the presence of electric (magnetic) field  can makes unstable the CRM against radial perturbations (Fig. \ref{fig:tmoencon}$(a)$). Thus the CRM cannot apply for $b=3$ (dotted curve) . The remaining functions are plotted for the value of the parameters representing a  physically acceptable CRM (Fig. \ref{fig:tmoencon}$(b)$ and \ref{fig:tdenelecmag}).  Also these have a similar  behavior  to the previous case.

\subsection{CRM for Kerr like disks}

A Kerr-like solution to the Einstein-Maxwell equations is  
\begin{subequations}
\begin{eqnarray}
\Phi    &=& \ln \left [ \frac {a^2x^2-b^2y^2-1}{(ax+1)^2-b^2y^2} \right ], \\
        & & \nonumber   \\
\Lambda &=& 2 \ln \left [ \frac {a^2x^2-b^2y^2-1}{a^2(x^2-y^2)}  \right ], \\
        & & \nonumber   \\
\psi    &=& \frac {2 \sqrt 2 by}{(ax+1)^2-b^2y^2}, 
\end{eqnarray}\label{eq:reis}
\end{subequations}
with  $a^2-b^2=1$. $x$ and $y$ are, again, the prolate spheroidal coordinates, given by Eqs. (\ref{eq:cooprox}) and (\ref{eq:cooproy}). This solution can  be generated, in these coordinates,  using a well-know theorem  proposed by Bonnor \cite{Bon1, Bon2} from  Kerr solution. The previous solution can also be obtained using this same theorem from  Taub-NUT solution. For $b=0$, it also goes over into the Darmois solution.  Thus, again,  $b$ is the  electric parameter. One can also obtain its  magnetostatic counterpart  cumputing the magnetostatic potencial using (\ref{eq:potmag}), and gives
\begin{equation}
A = -\frac{\sqrt 2 kb(1-y^2)(ax+1)}{a(a^2x^2-b^2y^2-1)},  
\end{equation}
again with  $a^2-b^2=1$. This solution describes the field of a massive magnetic dipole.  The physical quantities associated with disk now can be written as 
\begin{equation}
\begin{array}{ccl}
\tilde {\epsilon} & = & 8a^4\bar{y}\{ (\bar{x}^2-\bar{y}^2)[a(\bar{x}^2-1)[(a\bar{x}+1)^2+b^2\bar{y}^2] -2b^2\bar{x}(a\bar{x}+1)(1-\bar{y}^2)]  \\  
& &  -2\bar{x}(\bar{x}^2-1)(1-\bar{y}^2)[(a\bar{x}+1)^2-b^2\bar{y}^2]\}  \\
& &   /(a^2\bar{x}^2-b^2\bar{y}^2-1)^2[(a\bar{x}+1)^2-b^2\bar{y}^2]^2,
\end{array}
\end{equation}

\begin{equation}
\tilde {p}_\varphi = \frac {16a^4\bar{x}\bar{y}(\bar{x}^2-1)(1-\bar{y}^2)}{(a^2\bar{x}^2-b^2\bar{y}^2-1)^2[(a\bar{x}+1)^2-b^2\bar{y}^2]},
\end{equation}

\begin{equation}
\tilde{\rm j}_t = -\frac {4 \sqrt 2 a^4b(\bar{x}^2-\bar{y}^2)\{\bar{x}(1-\bar{y}^2)[(a\bar{x}+1)^2+b^2\bar{y}^2]-2a\bar{y}^2(a\bar{x}+1)(\bar{x}^2-1)\}  }{(a^2\bar{x}^2-b^2\bar{y}^2-1)[(a\bar{x}+1)^2-b^2\bar{y}^2]^3},   
\end{equation}

\begin{equation}
{\rm j}_\varphi = -\frac{2 \sqrt 2 a^4b\bar{y}(\bar{x}^2-1)(1-\bar{y}^2)(\bar{x}^2-\bar{y}^2)[(a\bar{x}+1)(3a\bar{x}+1)+b^2\bar{y}^2]}{(a^2\bar{x}^2-b^2\bar{y}^2-1)^3[(a\bar{x}+1)^2-b^2\bar{y}^2]},
\end{equation}
where $\bar{x}$ and  $\bar{y}$ are given by Eqs. (\ref{eq:xbar}) and (\ref{eq:ybar}).

\begin{figure}
$$
\begin{array}{cc}
\tilde {\epsilon} &  \tilde {p}_\varphi \\
\epsfig{width=3in,file=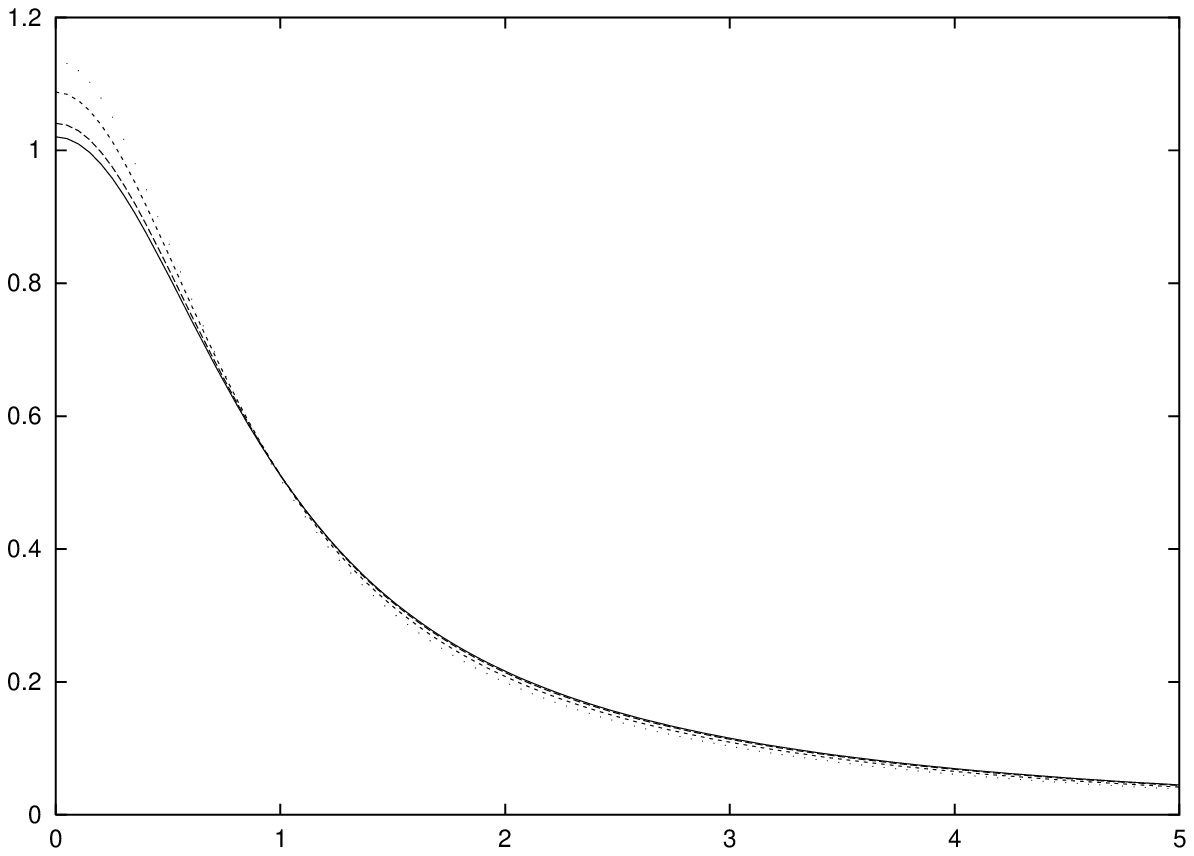} & \epsfig{width=3in,file=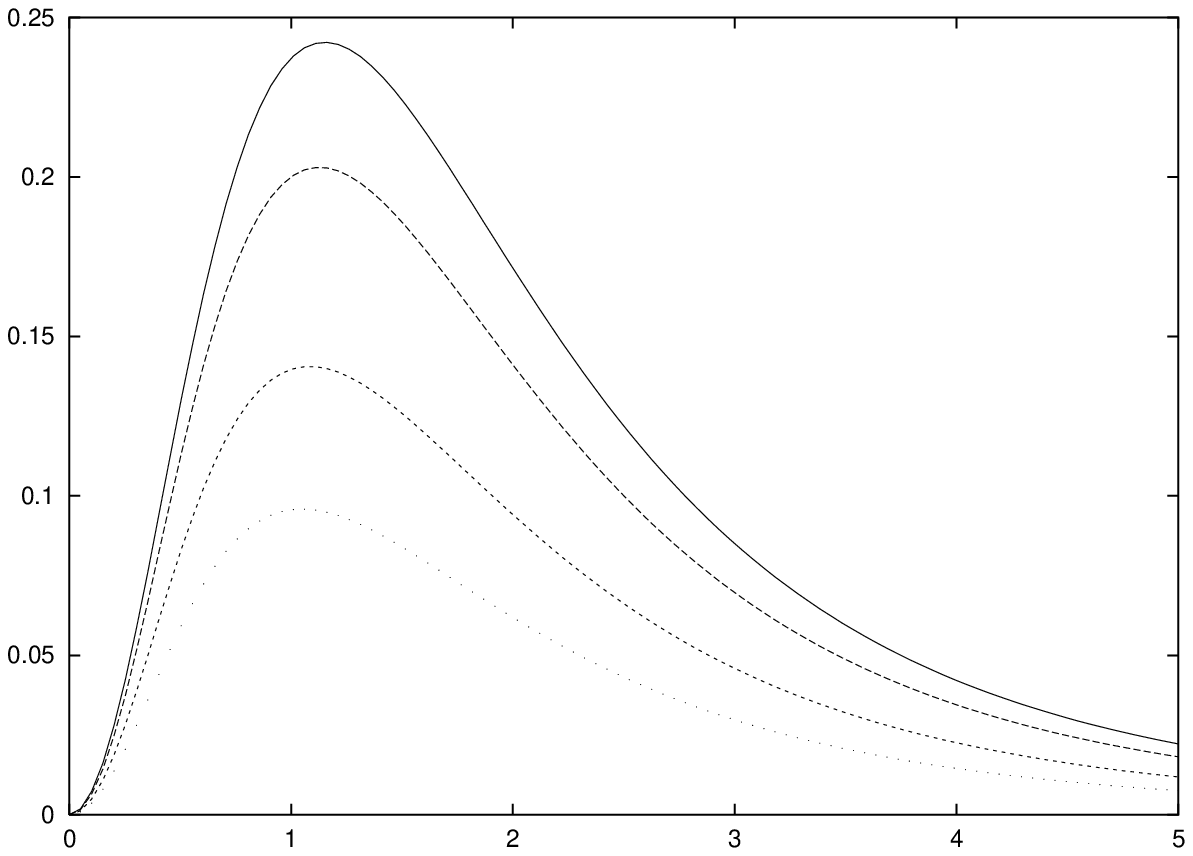}   \\
\tilde r  & \tilde r   \\
(a)  &  (b)      \\
\end{array}
$$	
\caption{$(a)$ Energy density  $\tilde {\epsilon}$ and $(b)$ azimuthal pressure  $\tilde {p}_\varphi$ as functions of $\tilde{r}$ for disks  with $\alpha=1.8$ and $b=0$ (solid curve), $0.5$, $1.0$, and $1.5$ (dotted curve).}\label{fig:kenerpres}
\end{figure}

\begin{figure}
$$
\begin{array}{cc}
\tilde {\rm j}_t &  \rm j_\varphi    \\
\epsfig{width=3in,file=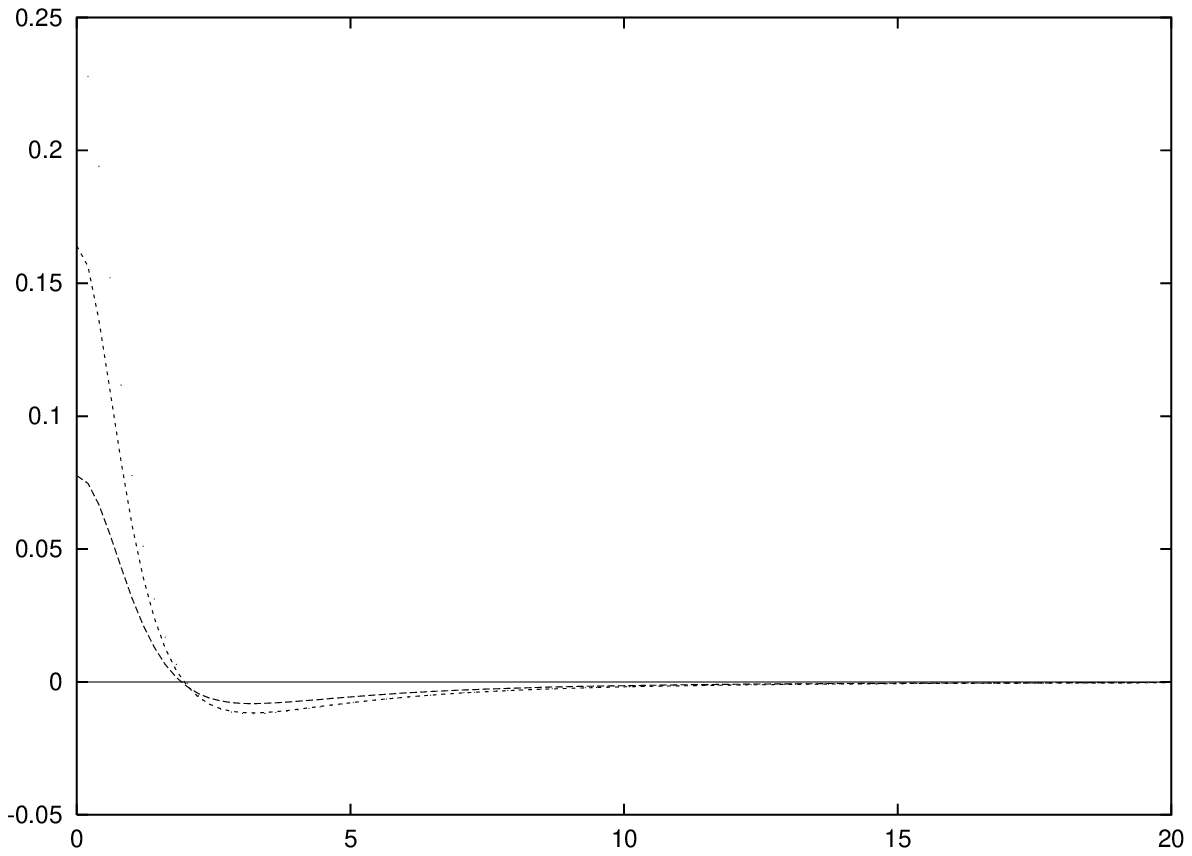} & \epsfig{width=3in,file=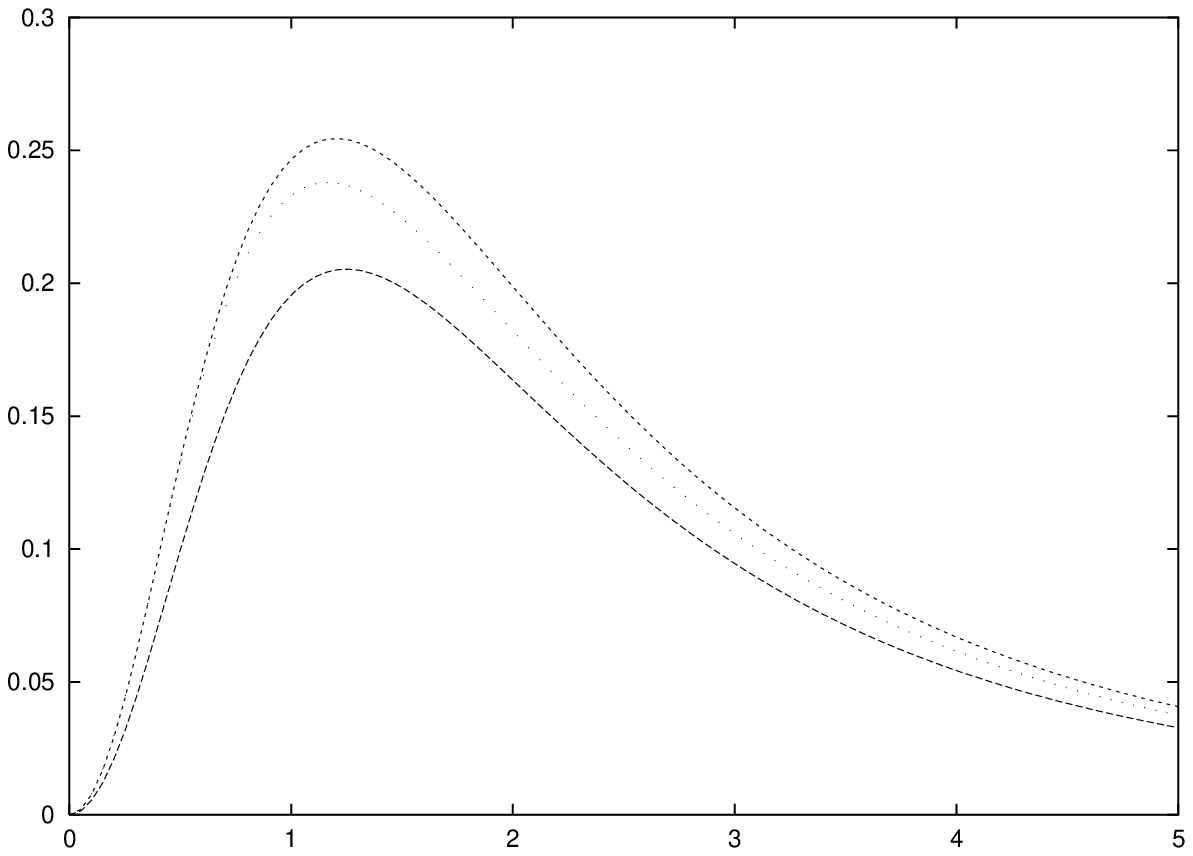}  \\
\tilde r  & \tilde r   \\
(a)  &  (b)   
\end{array} 
$$	
\caption{Current density: $(a)$ $\tilde {\rm j}_t$ and $(b)$  ${\rm j}_\varphi$ as functions of $\tilde r$ for disks  with $\alpha=1.8$ and $b=0$ (axis  $\tilde r$), $0.5$, $1.0$, and $1.5$ (dotted curve).}\label{fig:kcorelecmag}
\end{figure}

\begin{figure}
$$
\begin{array}{cc}
\rm U^2  & \rm U^2   \\
\epsfig{width=3in,file=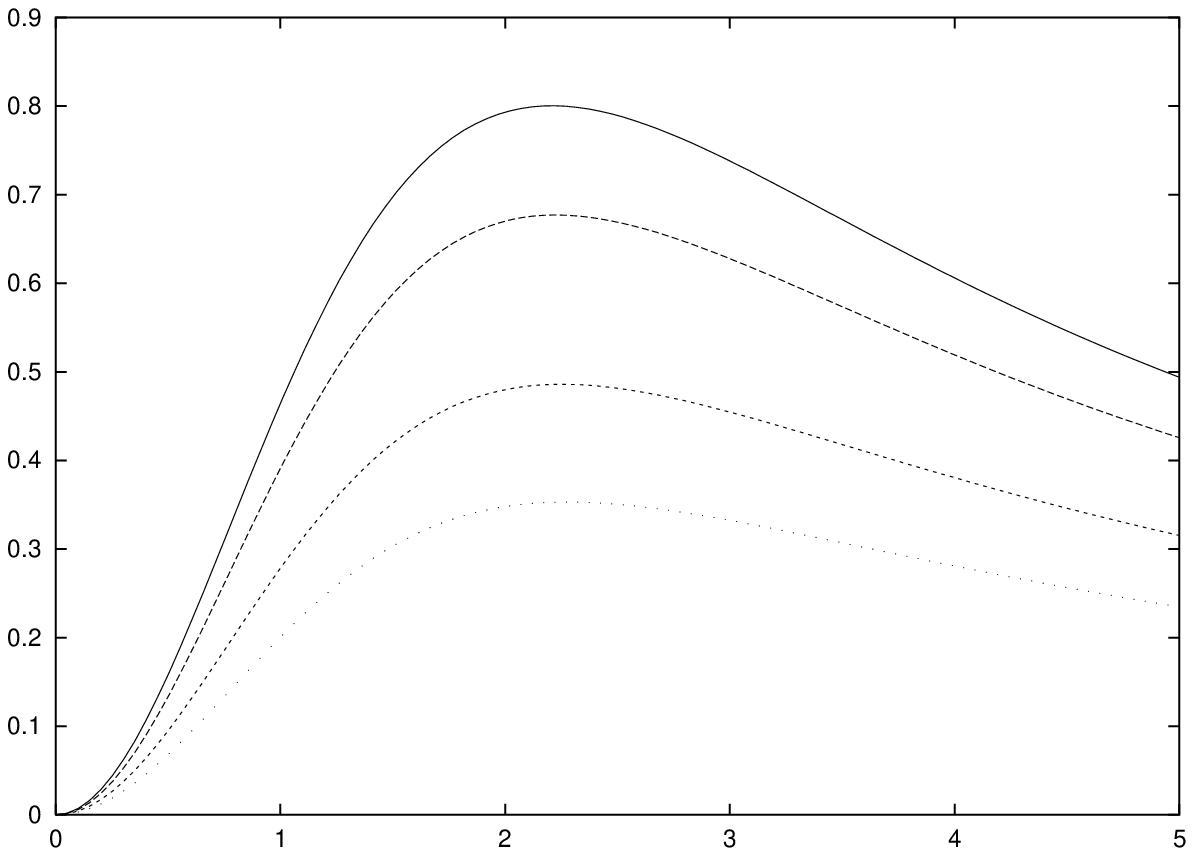} &  \epsfig{width=3in,file=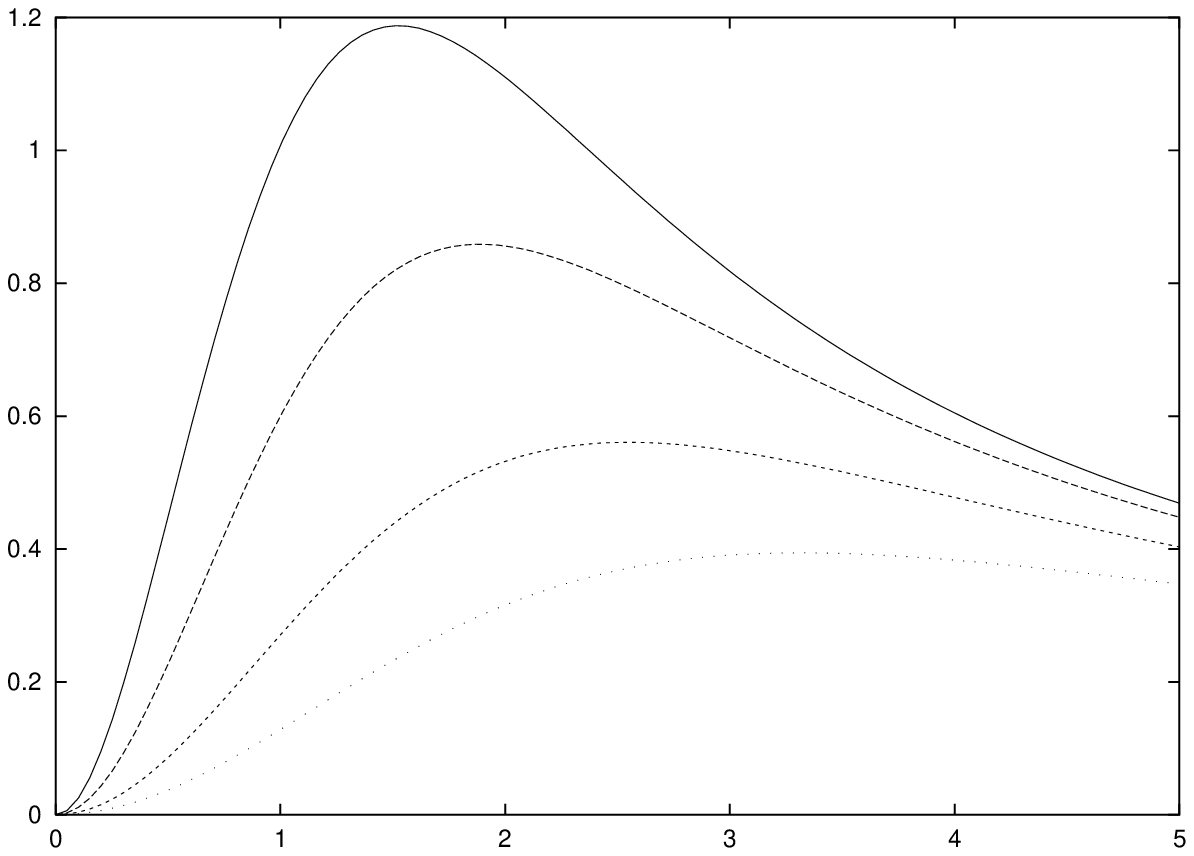} \\
\tilde r  & \tilde r \\
(a)  & (b)
\end{array}
$$	
\caption{ Tangential velocity $\rm U ^2$   as function of $\tilde r$ for disks  with $(a)$ $\alpha=1.8$ and $b=0$ (solid curve), $0.5$, $1.0$, and $1.5$ (dotted curve), and $(b)$  $b = 0.5$ and $\alpha = 1.4$ (solid curve), $1.6$, $2.0$, and $2.5$ (dotted curve)} \label{fig:kvel}
\end{figure}

\begin{figure}
$$
\begin{array}{cc}
\tilde h^2 &  \tilde \epsilon _\pm   \\
\epsfig{width=3in,file=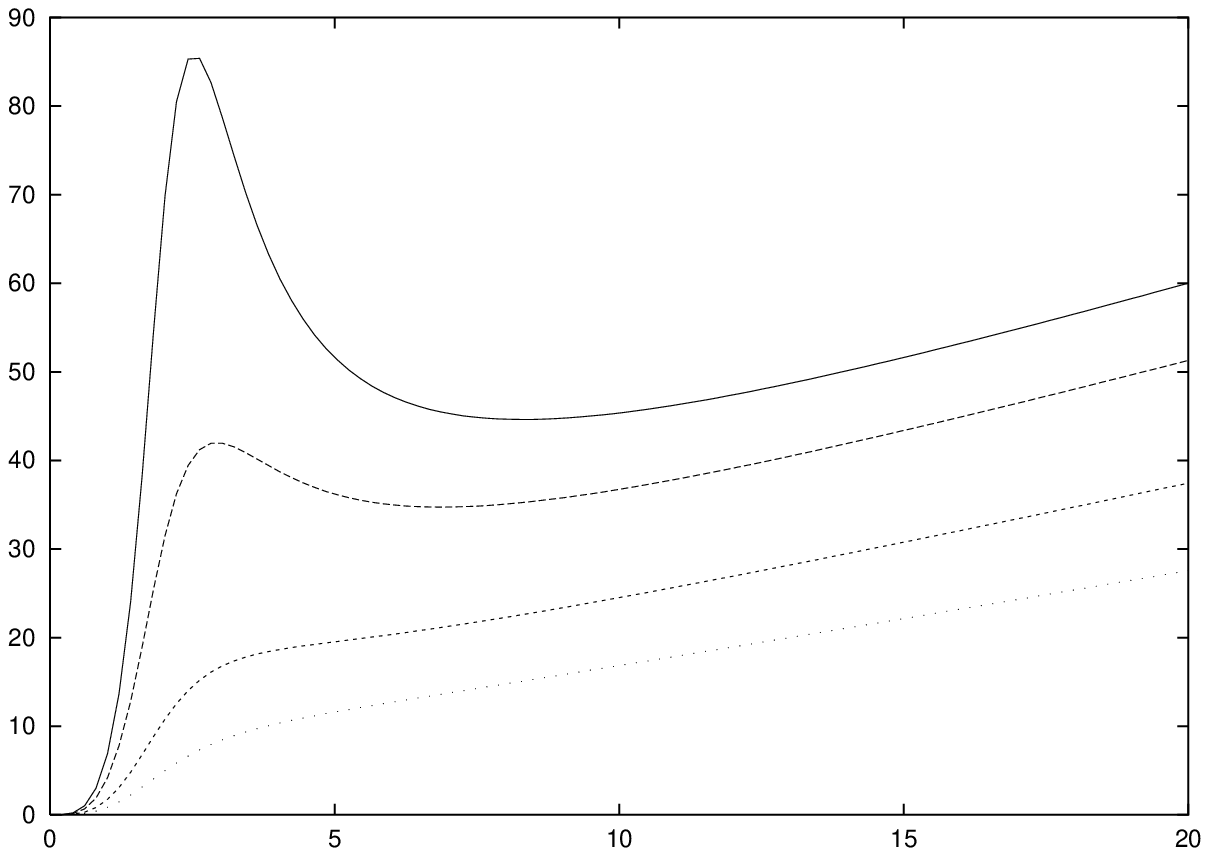} & \epsfig{width=3in,file=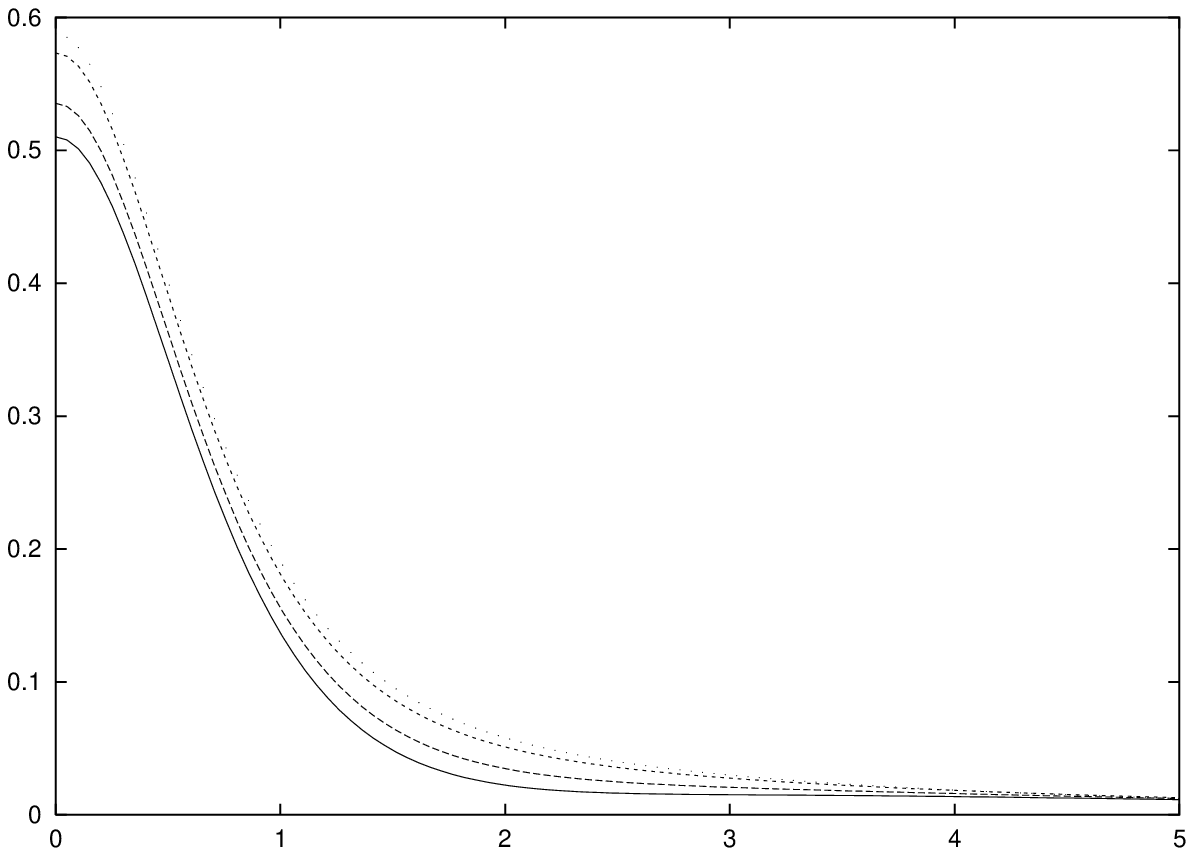} \\
\tilde r & \tilde r \\
(a) & (b)       
\end{array}
$$	
\caption{$(a)$ Specific angular momentum $\tilde h^2$  and $(b)$ mass densities  $\tilde \epsilon _\pm$ as functions of $\tilde r$ for disks  with $\alpha=1.8$ and $b=0$ (solid curve), $0.5$, $1.0$, and $1.5$ (dotted curve). }\label{fig:kmoencon}
\end{figure}

\begin{figure}
$$
\begin{array}{cc}
- \tilde \sigma _{e \pm} &  \tilde \sigma _{m \pm}    \\
\epsfig{width=3in,file=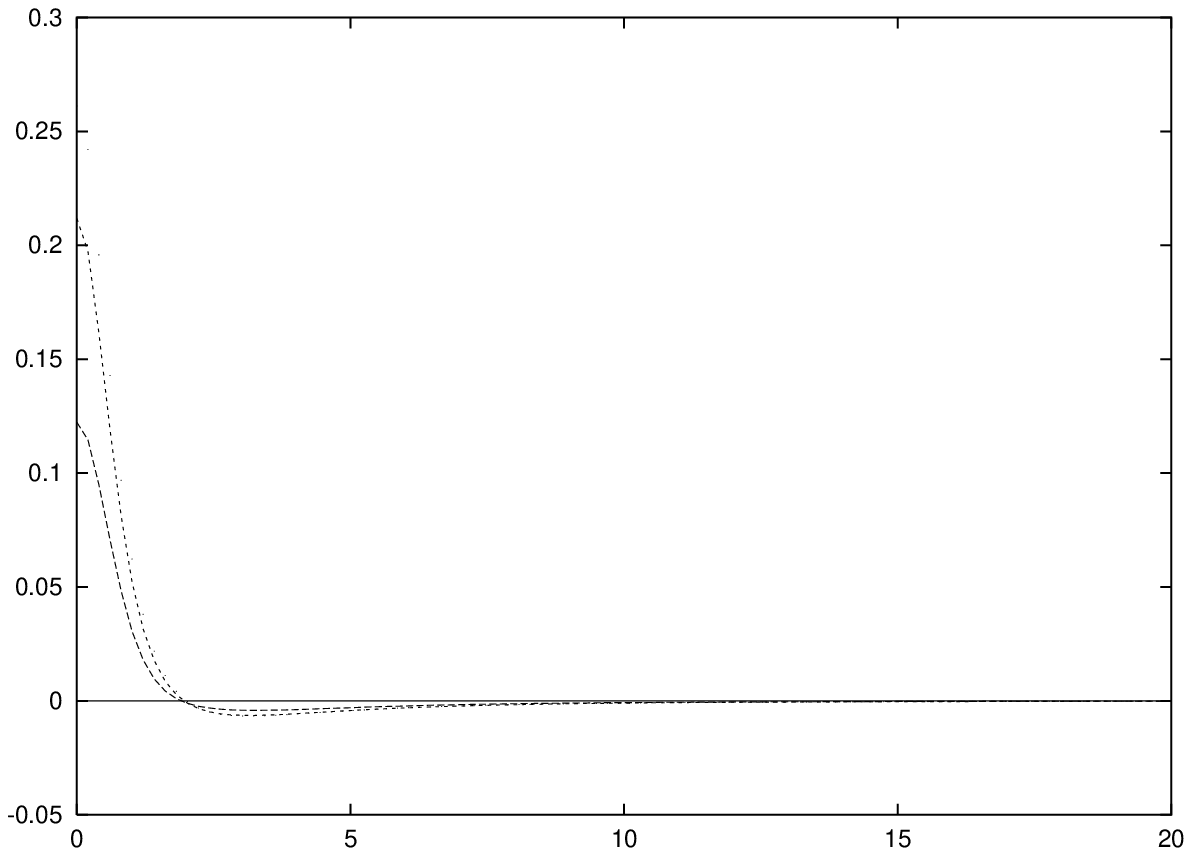} & \epsfig{width=3in,file=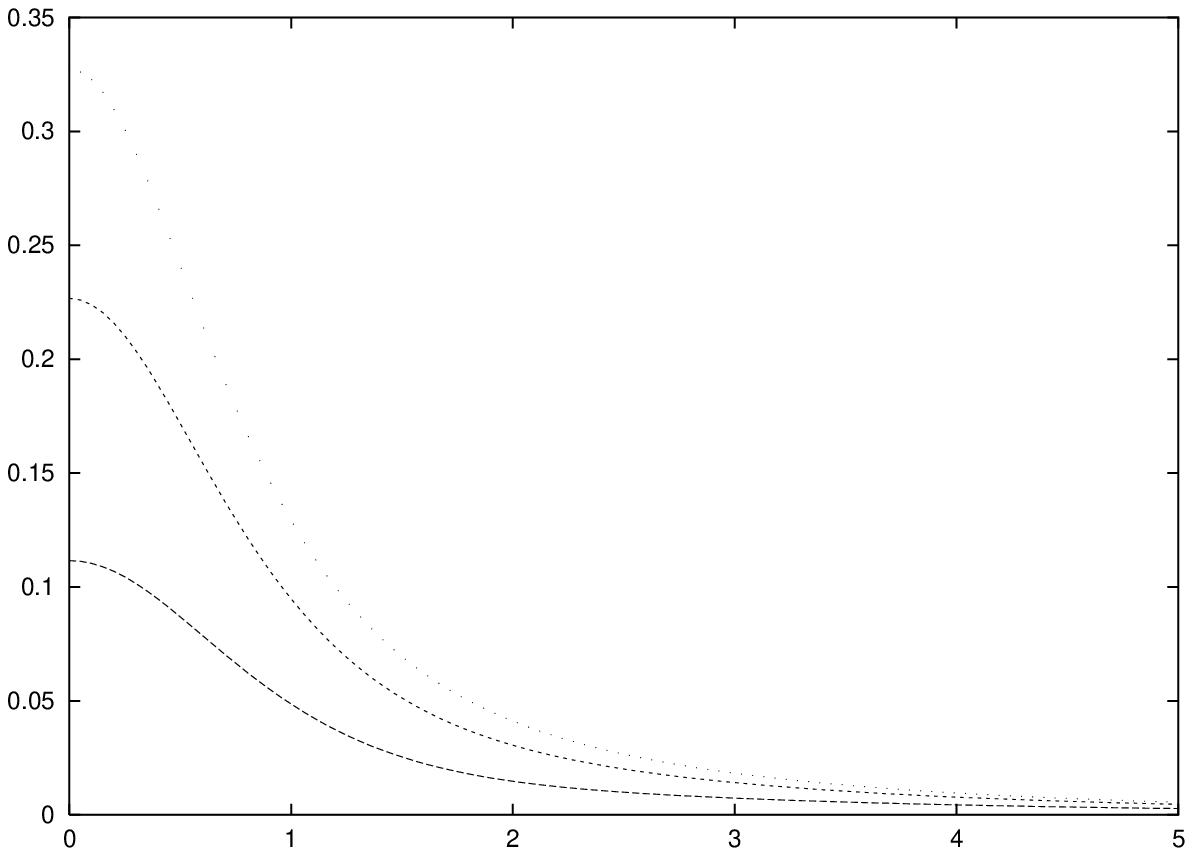} \\
\tilde r & \tilde r \\
(a)   &  (b) 
\end{array}
$$	
\caption{Charge densities: $(a)$ $\tilde \sigma _{e \pm}$ and $(b) $$\tilde \sigma _{m \pm}$ as functions of $\tilde r$ for disks  with $\alpha=1.8$ and $b=0$ (axis $\tilde r$), $0.5$, $1.0$, and $1.5$ (dotted curve).}\label{fig:kdenelecmag}
\end{figure}

In figs. \ref{fig:kenerpres} and \ref{fig:kcorelecmag} the plots of the physical quantities describing the disks  are shown for $\alpha=1.8$ and $b=0$, $0.5$, $1.0$, and $1.5$. The energy density  behaves in the opposite way to the previous cases. That is, near to the disk center it increases when   the electric (magnetic) field is applied and decreases later. The other quantities have a  similar behavior to the precedent cases. However, the electrostatic current after some r  takes  negative values. 
Likewise, the quantities corresponding to the CRM are shown in follo\-wing figures for the same value of the of the parameters $\alpha$ and $b$. Theses CRM are more relativistic than   the ones above  considered (Fig. \ref{fig:kvel}(a)). Note  that (Fig. \ref{fig:kvel}(b)) the disks with $b=0.5$ and $\alpha = 1.4 $ (solid curve) cannot be constructed from the CRM because $\rm{U}^2>1$.
We also note that  the presence of electric (magnetic) field  can stabilize the CRM against radial perturbations (Fig. \ref{fig:kmoencon}$(a)$). Therefore, only the CRM constructed with  $b=1.0$, and $1.5$ are well behaved. Finally, the other functions are drawn for the same value of the parameters (Figs. \ref{fig:kmoencon}$(b)$ and \ref{fig:kdenelecmag}).

\section{Discussion}

A detailed study of the Counter-Rotating Model for
generic  electrostatic  (magnetos\-tatic) axially symmetric thin disks without radial pressure was presented. A general constraint over the counter-rotating tangential
velocities was found, needed to cast the surface energy-momentum tensor
of the disk in such a way that it can be interpreted as the superposition
of two counter-rotating  charged dust fluids. The constraint found is
completely equivalent to the necessary and sufficient condition obtained
in reference \cite{FMP}.  We next showed that this constraint is satisfied if we take the two counter-rotating  fluids as circulating along electrogeodesics with  equal and opposite tangential velocities.
We also have obtained explicit expressions for the energy
densities, electrostatic (magnetostatic) current densities and velocities   of the counter-rotating streams in terms of the
energy density, azimuthal pressure and planar current density of the disk, that are also equivalent to the correspondig expressions in reference \cite{FMP}.

Three specific examples were considered in the present work based in simple solutions to the Einsteins-Maxwell equations generated by conventional solution-generating tecniques \cite{KSHM}. We found that the CRM 
for Kerr-like disks  are more relativistics than the ones obtained from Taub-NUT and Reissner-Nordstr\"{o}m  like solutions.
We also  saw that  the   presence of electric (magnetic) field  can make unstable  the CRM against radial perturbations in the case of Taub-NUT and Reissner-Nordstr\"{o}m  like disks, and conversely, stabilize the CRM in the case of Kerr-like disks. We also constructed some  CRM  with well defined counter-rotating tangential velocities and stable against radial perturbations. 
 
On the other hand, the generalization of the Counter-Rotating
Model presented here to the case  with radial pressure is in consideration. Also, the generalization to  rotating thin disks with or without
radial pressure  in presence of electromagnetic fields is being considered.
   
\subsection*{Acknowledgments}

Gonzalo Garc\'\i a R. wants to thank a Fellowship from Vicerrector\'\i a Acad\'emica,
Universidad Industrial de Santander.

\newpage


\begin{thebibliography}{9999}

\bibitem{BS} W. A. Bonnor and A. Sackfield, Comm. Math. Phys. 8, 338
(1968).

\bibitem{MM1} T. Morgan and L. Morgan, Phys. Rev. 183, 1097 (1969).

\bibitem{MM2} L. Morgan and T. Morgan, Phys. Rev. D2, 2756 (1970).

\bibitem{CHGS} A. Chamorro, R. Gregory and J. M. Stewart,
Proc. R. Soc. Lond. A 413, 251 (1987).

\bibitem{GL1} G. A. Gonz\'alez and P. S. Letelier. Class. Quantum. Grav.
16, 479 (1999).

\bibitem{LP} D. Lynden-Bell and S. Pineault, Mon. Not. R. Astron. Soc.
185, 679 (1978). \label{bib:LP}

\bibitem{LO} P.S. Letelier and S. R. Oliveira, J. Math. Phys. 28, 165
(1987).

\bibitem{LEM} J. P. S. Lemos, Class. Quantum Grav. 6, 1219 (1989).

\bibitem{LL1} J. P. S. Lemos and P. S. Letelier, Class. Quantum Grav. 10,
L75 (1993).

\bibitem{BLK} J. Bi\u{c}\'{a}k, D. Lynden-Bell and J. Katz, Phys. Rev.
D47, 4334 (1993).

\bibitem{BLP} J. Bi\u{c}\'{a}k, D. Lynden-Bell and C. Pichon, Mon. Not.
R. Astron. Soc. 265, 126 (1993).

\bibitem{BL} J. Bi\u{c}\'ak and T. Ledvinka. Phys. Rev. Lett. 71, 1669
(1993).

\bibitem{LL2} J. P. S. Lemos and P. S. Letelier, Phys. Rev D49, 5135
(1994).

\bibitem{LL3} J. P. S. Lemos and P. S. Letelier, Int. J. Mod. Phys. D5,
53 (1996).

\bibitem{KLE} C. Klein, Class. Quantum Grav. 14, 2267 (1997).

\bibitem{GL2} G. A. Gonz\'alez and P. S. Letelier. Phys. Rev. D 62,
064025 (2000). \label{bib:GL2}

\bibitem{RGK} V. C. Rubin, J. A. Graham and J. D. P Kenney. Ap. J. 394,
L9-L12, (1992).

\bibitem{RFF} H. Rix, M. Franx, D. Fisher and G. Illingworth. Ap. J. 400,
L5-L8, (1992).

\bibitem{LET1} P. S. Letelier. Phys. Rev. D 60, 104042 (1999). 

\bibitem{LBZ} T. Ledvinka, J. Bi\u{c}\'{a}k, and M. \u{Z}ofka, in {\it Procceeding of 8th Marcel-Grossmann Meeting in General Relativity}, edited by T. Piran  (Worls Scientific, Singapure, 1999)

\bibitem{KBL} J. Katz, J. Bi\u{c}\'ak and D. Lynden-Bell, Class. Quantum
Grav. 16, 4023 (1999).

\bibitem{LET2} P. S. Letelier. Phys. Rev. D 22, 807 (1980).

\bibitem{FMP} J. J. Ferrando, J. A. Morales and M. Portilla. Gen. Rel. and
Grav. 22, 1021 (1990).

\bibitem{CHAN} S. Chandrasekar, {\it The Mathematical Theory of Black
Holes}. (Oxford University Press, 1992).

\bibitem{FLU} L. D. Landau and E. M. Lifshitz, {\it Fluid Mechanics} (Addisson-Wesley, Reading, MA, 1989).

\bibitem{KSHM} D. Kramer, H. Stephani, E. Herlt, and  M. McCallum, {\it Exact Solutions of Einsteins's  Field Equations} (Cambridge University Press, Cambridge, England, 1980).
  
\bibitem{E1} F. J. Ernst. Phys. Rev. D 167, 1175 (1968).

\bibitem{E2} F. J. Ernst. Phys. Rev. D 168, 1415 (1968).

\bibitem{Sch} K.  Schwarzschild. Sitzungsber. Preuss. Akad. Wiss., 189 (1968).

\bibitem{Bon1} W. B. Bonnor. Z. Phys. 161, 439  (1961).

\bibitem{Bon2} W. B. Bonnor. Z. Phys. 190, 444  (1966).

\end{thebibliography}
\end{document}